\documentclass[10pt,twocolumn,superscriptaddress,floatfix]{revtex4-2}

\usepackage[utf8]{inputenc}
\usepackage[T1]{fontenc}
\usepackage[colorlinks=true,linkcolor=blue,urlcolor=blue,citecolor=blue,breaklinks=true]{hyperref}

\usepackage{amsmath,amsthm,amssymb,mathtools} 

\usepackage{siunitx}
\DeclareSIUnit\fps{fps}

\usepackage{graphicx}

\usepackage{color}
\definecolor{blue}{rgb}{0,0,1}

\usepackage{makecell}
\usepackage{booktabs}
\usepackage{multirow}

\begin{document}

\title{Microscopic fluctuations in power-grid frequency recordings at the sub-second scale}

\author{Benjamin~Schäfer}
\thanks{contributed equally}
\affiliation{Institute for Automation and Applied Informatics, Karlsruhe Institute of Technology, 76344 Eggenstein-Leopoldshafen, Germany}

\author{Leonardo~Rydin~Gorj\~ao}
\thanks{contributed equally}
\affiliation{Department of Computer Science, OsloMet -- Oslo Metropolitan University, N-0130 Oslo, Norway}
\affiliation{Forschungszentrum J\"ulich, Institute for Energy and Climate Research - Systems Analysis and Technology Evaluation (IEK-STE), 52428 J\"ulich, Germany}

\author{G.~Cigdem~Yalcin}\thanks{Correspondence: benjamin.schaefer@kit.edu and gcyalcin@istanbul.edu.tr}
\affiliation{Department of Physics, Istanbul University, 34134, Vezneciler, Turkey} 

\author{Ellen~Förstner}
\affiliation{Institute for Automation and Applied Informatics, Karlsruhe Institute of Technology, 76344 Eggenstein-Leopoldshafen, Germany}

\author{Richard~Jumar}
\affiliation{Institute for Automation and Applied Informatics, Karlsruhe Institute of Technology, 76344 Eggenstein-Leopoldshafen, Germany}

\author{Heiko Maass}
\affiliation{Institute for Automation and Applied Informatics, Karlsruhe Institute of Technology, 76344 Eggenstein-Leopoldshafen, Germany}

\author{Uwe~Kühnapfel}
\affiliation{Institute for Automation and Applied Informatics, Karlsruhe Institute of Technology, 76344 Eggenstein-Leopoldshafen, Germany}

\author{Veit~Hagenmeyer}
\affiliation{Institute for Automation and Applied Informatics, Karlsruhe Institute of Technology, 76344 Eggenstein-Leopoldshafen, Germany}

\begin{abstract}
Complex systems, such as the power grid, are essential for our daily lives. Many complex systems display (multi-)fractal behavior, correlated fluctuations and power laws. Whether the power-grid frequency, an indicator about the balance on supply and demand in the electricity grid, also displays such complexity remains a mostly open question. Within the present article, we utilize highly resolved measurements to quantify the properties of the power-grid frequency. We show that below 1 second, the dynamics may fundamentally change, including a suddenly increasing power spectral density, emergence of multifractality and a change of correlation behavior. We provide a simplified stochastic model involving positively correlated noise to reproduce the observed dynamics, possibly linked to frequency dependent loads. Finally, we stress the need for high-quality measurements and discuss how we obtained the data analyzed here.
\end{abstract}

\maketitle

\section{Introduction}
Complex systems, sometimes natural, sometimes man-made, surround us at all scales as an integral part of everyday life, ranging from neural networks at small scales to global pandemics or the world wide web at large scales~\cite{newman2003structure}.
To satisfy most basic needs of modern people in every day life, we rely on power systems, which are an important example of a complex system~\cite{arenas2008synchronization,pagani2013power}.
This dependence ranges from the simple cooling of food, vital medical care, long-distance transportation, instant communication, industrial automation, to advanced scientific experiments, many of which are impossible without electricity.
Since reliability and stability are essential for our power grid, analysis of power systems as a complex system has become critically important~\cite{witthaut2022collective}.

A distinguishing feature of complex systems is the emergence of new phenomena and complex dynamics based on the various nonlinear interactions of a large collection of components.
One particularly interesting and universal property is the observation of power laws~\cite{newman2005power}.
Under power laws we understand that a function decays as $f(x) \sim x^{-\beta}$, for large $|x|$, with some constant $\beta>0$.
Such power laws have been investigated in real-world data sets from a wide variety of fields, including biological and technical systems~\cite{Clauset2009}.
Empirically, they are often discussed as the emergence of heavy or fat tails, e.g. web nodes with an extremely large number of connections exist more often than intuitively expected.
From a theoretical point of view, the scale-invariance and universality of power-laws explains their popularity in many models~\cite{albert2000topology}.
Furthermore, the fractal dimension of a system can be used to investigate dynamic processes in complex systems and is indeed one of the most basic and important quantities to characterize a system~\cite{wu2019fractal}.
It captures how a system with many degrees of freedom effectively only occupies a fraction of the high-dimensional state space.

In the present paper, we focus on power grids, i.e. electrical supply networks and their dynamics as a complex system.
While the network itself easily constitutes a complex system~\cite{witthaut2022collective}, less attention has been paid to individual local dynamics~\cite{anvari2020stochastic}.
One key quantity for control and monitoring purposes of the power system is the power-grid frequency.
It follows the balance of generation and consumption and allows insights into the general stability of the grid as well as the impact of renewable generators~\cite{machowski2020power, schafer2018non,han2022collective}.
So far only few studies have investigated power laws or even (multi-)fractality in power grids~\cite{zhou2020fractal}, partially because data are not easily available.

The dynamics of the power-grid frequency is hard to describe analytically and even numerically, due to the non-linear interactions of power generators and consumers, the complex network topology, the large number of individuals as well as companies, countries, and added stochastic effects, e.g. stemming from renewable generation~\cite{anvari2020stochastic,kruse2021revealing}.
This stresses the need to acquire high-quality and trustworthy data on power systems as an approach to quantitatively describe and understand the power grid as a complex system.
Unfortunately, reliable, high-quality continental and global measurements from grid frequency are only rarely made publicly available~\cite{rydin2020open}.

To obtain high-quality measurements, initiatives such as Gridradar~\cite{Gridradar} and GridEye/FNET~\cite{chai_wide-area_2016} have been developed.
From the academic side, the Electrical Data Recorder (EDR) has been developed as a device for high resolution time series acquisition in low voltage distribution girds.
It provides UTC-time-stamped frequency estimates up to ten times per second and allows long-term storage of full voltage and/or current wave forms for further analysis~\cite{maass2013first,maas2015data, jumar2021database,foerstner2022experimental}.
The frequency is estimated using the well-known zero-crossing technique that enables straightforward and fast processing and is widely developed by scientists~\cite{zhao2020recent, grillo2012efficient, mendonca2014least, luo2018improved}.

Within the present article, we show how the frequency displays complex behavior on time scales below one second.
In particular, we observe an increasing spectral density, emerging multifractality and a switch from correlated to anti-correlated time series.
We further highlight differences in short-term fluctuations among various power grids.
Finally, we emphasize the need for trustworthy, highly resolved measurements of the power-grid frequency and show one way of how these can be obtained.

\section{Results} 
\begin{figure}[t]
 \includegraphics[width=\linewidth]{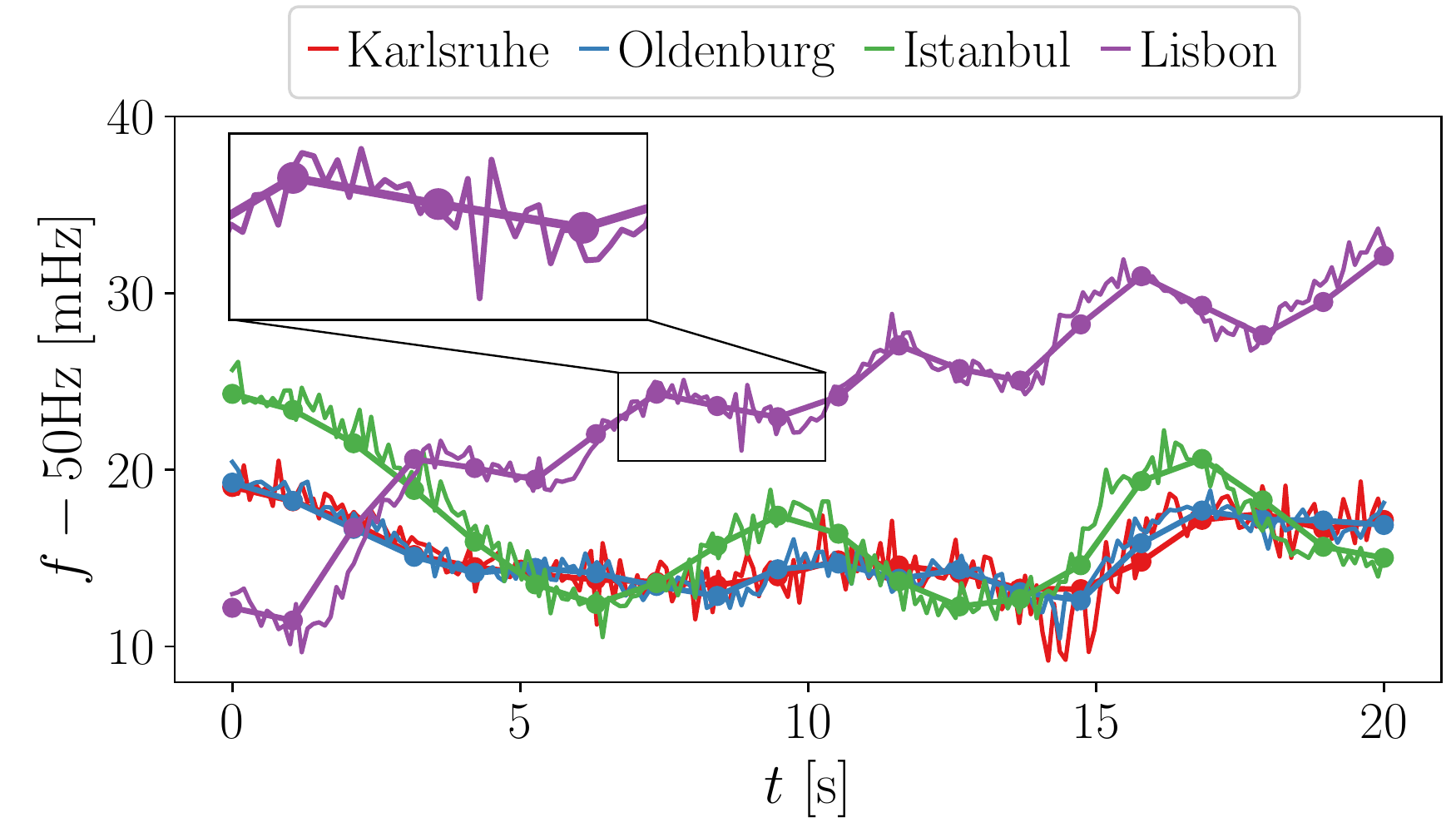}  
 \caption{\textbf{Snippets of power-grid frequency trajectories.}
 power-grid frequency trajectories from 2019-07-27~06:00 in Continental Europe, for 20 seconds, including 1 second resolution and \SI{100}{\milli\second} resolution data.
 The smaller \SI{100}{\milli\second} resolution recordings fluctuate strongly around the \SI{1}{\second} resolution data, sometimes with very large jumps.
 The frequency dynamics is revealed as fractal for this exemplary trajectory: If we ``zoom in'' to higher temporal resolutions, we observe that the data is self-affine, i.e. the data displays self-similarity in its fluctuations.
 In the inset we focus on a 2-second span, where we can observe both small fluctuations (in the 1st second) and large fluctuations (in the 2nd second).
 }\label{fig:trajectories}
\end{figure}

To analyze empirical complex systems, such as the power-grid frequency dynamics, we require data.
Some measurements with one second resolution are readily available from transmission system operators (TSOs) for analysis and comparison with independent measurements~\cite{rydin2020open}.
Meanwhile, time series with higher temporal resolution are not easy to obtain.
Still, this higher temporal resolution is critical to uncover key properties of the underlying system, see e.g.~\cite{RydinGorjao2021a,RydinGorjao2021b} and Fig.~\ref{fig:trajectories}: Representative frequency trajectories on \SI{1}{\second} and on \SI{100}{\milli\second} (\SI{0.1}{\second}) resolutions clearly display different behavior, with notable fluctuations and high volatility on the short time scale.

We address observable changes in the spectrum towards short time scales as well as the correlation and fractal behavior of the time series.
To understand these observations and ensure a high data quality, we outline how the recordings have been obtained.
We therefore describe the EDR measurement process and focus on its reliability.

\subsection{Emerging phenomena on short time scales}
The following observations are based on newly available \SI{100}{\milli\second} frequency estimates and particularly focus on phenomena emerging at the newly-accessible time scale below one second.
The acquisition of the high-resolution frequency data is described later in subsection~\ref{secsub:ObtainingFrequencyRecordings}.

\begin{figure*}[t]
 \includegraphics[width=\linewidth]{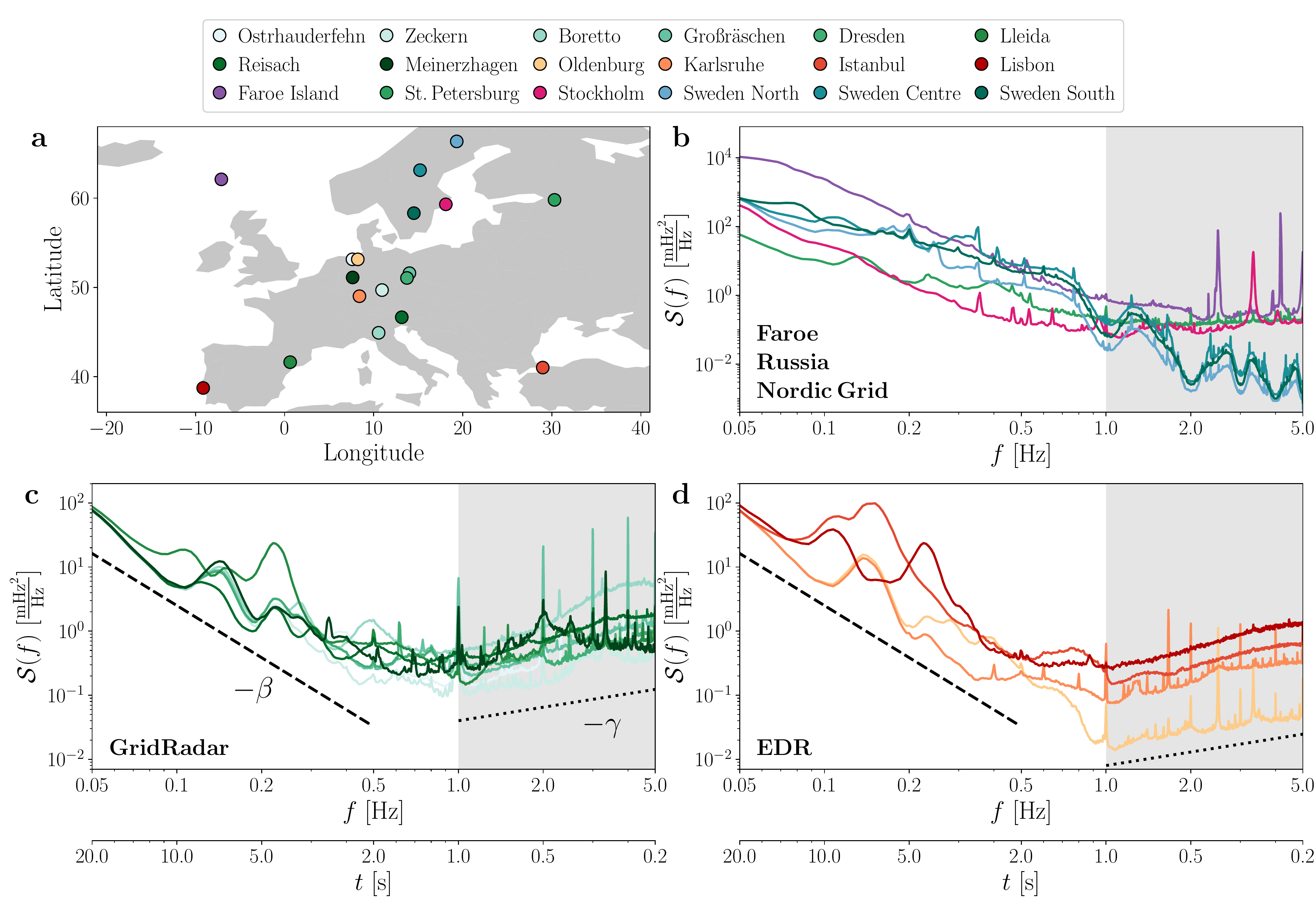}
 \caption{\textbf{Power grid dynamics change qualitatively at the 1-second time scale.} 
 We utilize recordings from various locations in Europe (\textbf{a}) to then plot the spectral density $\mathcal{S}(f)$ of the power-grid frequency (\textbf{b}-\textbf{d}).
 This analysis includes independent synchronous areas (\textbf{b}) as well as Continental Europe data obtained from Gridradar (\textbf{c}) and from the EDR (\textbf{d}). 
 Note that we plot $\mathcal{S}(f)$ and also report the corresponding time scales $t=1/f$.
 While all recordings initially display an approximately decaying power law $\beta$, several recordings display a rising spectrum with slope $\gamma$ for time scales below $t\sim 1s$.}\label{fig:spectrum}
\end{figure*}

As one key quantity, we compute the power spectral density (PSD) $\mathcal{S}(f)$, which allows us to analyze the power distribution of different oscillations in the components of the frequency recorded from the system in a stochastic process~\cite{spectral2021}.
As with many processes observed in nature, we also expect the power spectrum of self-affine signals to decay following a power-law $S(f)\sim f^{-\beta}$~\cite{krakovska2016fractal,mandelbrot1977}.
PSD is a common tool that allows us to perform further analysis on data originating from stochastic processes, such as data obtained from electroencephalogram (EEG)~\cite{eeg2020} as well as for power-grid frequency signal analysis, which we are currently considering here~\cite{spectral2021}.
Furthermore, we may also estimate other properties such as the fractality, i.e. self-similarity on various scales, or the correlation of the time series, which are closely connected to the spectrum~\cite{krakovska2016fractal,eke2002,mandelbrot1983}.

\subsubsection{Power-law analysis}
Computing the PSD for the power-grid frequency recordings reveals a new qualitative dynamics on the short time scale at essentially all measurement locations, see Fig.~\ref{fig:spectrum}.
We utilize several independent recordings (see Fig.~\ref{fig:spectrum}\textbf{a} for a map) from the Continental European grid (Fig.~\ref{fig:spectrum}\textbf{c} and \textbf{d}) and from three different synchronous areas Fig.~\ref{fig:spectrum}\textbf{b}.
Let us briefly remark on two observations we do not discuss in detail: 

First, on the longer time scale of $\sim 3...10$ seconds we observe pronounced inter-area oscillations, where different regions within one synchronous area oscillate against each other.
We notice the spatial dependency in the spectrum as different location show different magnitudes and position (in terms of time) of these peaks.
This is explained by the fact that e.g. North-South and East-West oscillations have a different frequency, see also~\cite{Klein1991,rydin2020open,RydinGorjao2021a} for a more detailed discussion of this phenomenon.

\begin{figure*}[t]
  \includegraphics[width=0.95\linewidth]{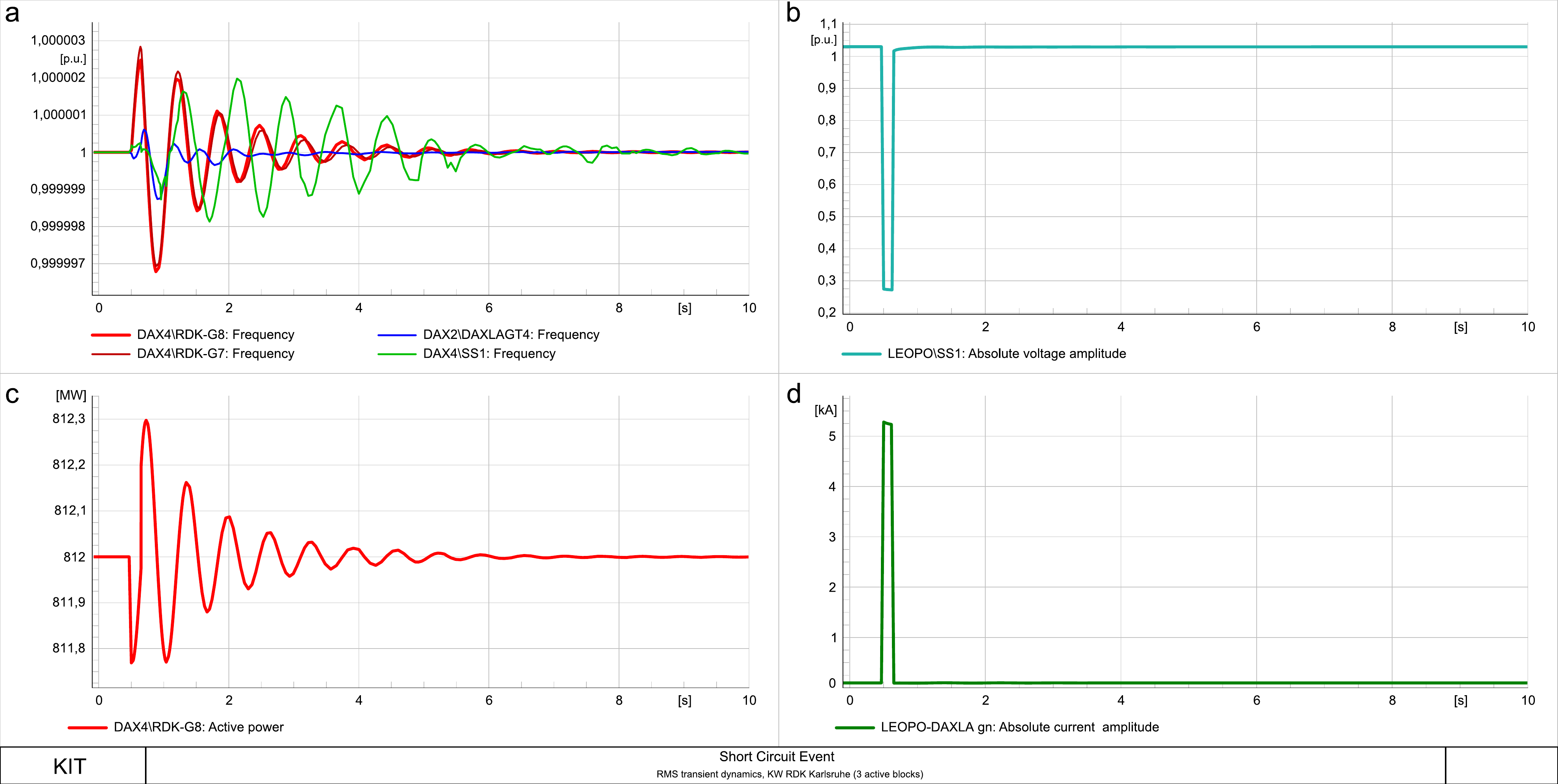}
  \caption{\textbf{Frequency and power dynamics in a transmission grid display characteristic sub-second dynamics.} We display the local frequencies at 380/220/110-kV station DAXLA with 3 powerstation-blocks (\textbf{a}), the real power of one block (\textbf{b}), as well as the absolute voltage amplitude (\textbf{c}) and the absolute current on the line that displayed a short-circuit to trigger the event (\textbf{d}). The transient dynamics is triggered via a three-phase short circuit between 0.5 and 0.65 seconds on a 110kV line between Daxlanden and Leopoldshafen (connection point of the KIT campus North). Line and node parameters use realistic values for the South-West German power grid. Frequency and voltage are given in the per-unit (p.u.) system, i.e. at 1 p.u. they are at the reference value.}
  \label{fig:network_simulation}
\end{figure*}

Second, there are various narrow and distinct peaks on the time scale below one second in the spectrum, see e.g. Fig.~\ref{fig:spectrum}\textbf{d} Karlsruhe and Oldenburg.
Curiously, not all recordings from the Continental European area display such peaks, e.g. neither Lisbon nor Istanbul.
Other synchronous areas again show mixed results: Recordings from Sweden only show less pronounced peaks, while the peaks are again present in the Faroe Island recordings.
These peaks could arise from characteristic eigenfrequencies of the power grid when subject to a disturbance, see Fig.~\ref{fig:network_simulation}.
While the short circuit with an active power step of approximately $\SI{440}{MW}$ only leads to a frequency change of the order $10^{-6}$ (0.15 mHz) in a realistic grid simulation (see Fig.~\ref{fig:network_simulation}a), 
the dynamics take place on the second to sub-second scale with eigenfrequencies of about \SI{1.5}{Hz}: Note the 3 full oscillations between 2 and 4 seconds in Fig.~\ref{fig:network_simulation}a and c.
Further investigations will be necessary to fully explore the origin and relevance of these peaks. In the current study, we instead focus on another key observation: The qualitative change in the spectrum around $1$ second and the emergence of complex phenomena at time scales $<1$ second (greyed out).

On the long time scale of $t>1$ second, all measurements display a power-law-like decay of the spectral density $\mathcal{S}(f) \sim f^{-\beta} \sim t^{\beta}$, with $f$ the frequency and $t=1/f$ the period.
For ease of interpretation we plot $\mathcal{S}(f)$ as a function of $f$ but also provide the corresponding $1/t$ values, see Fig.~\ref{fig:spectrum}.
This decay follows approximately $f^{-\beta}$, 
where we determine the exponent $-\beta$ as the slope of the dashed line in the double logarithmic plot for each measurement site separately.
As expected, the decay constant $\beta$ is positive, indicating a decrease of the spectral density for shorter periods (large frequencies) of the recordings.
This decreasing spectral density, i.e. the power-law, has been observed in numerous systems, ranging from many man-made systems, such as the world-wide-web over transport networks to biological systems~\cite{newman2005power, Clauset2009}.
Power-laws remain an active field of research, emerging both on microscopic and macroscopic scales.
It often remains open to which extend the individual units or their collective dynamics contribute to the observed power laws~\cite{bio2014}.

Meanwhile, the spectral density starts to increase on short time scales of $t<1$ second.
We again characterize the slope of the spectrum via a linear fit in the double-logarithmic plot and thereby determine the slope $\gamma$, which in this case is smaller than zero for all recordings from the Continental European power grid (Fig.~\ref{fig:spectrum}\textbf{c} and \textbf{d}).
Interestingly, this negative $\gamma$ is not exclusive to the Continental European region, see Fig.~\ref{fig:spectrum}\textbf{b}, where we show the power spectral density of five other power grid recordings, namely from the power grids on the Faroe Islands, Russia (St.~Petersburg) and Nordic (Sweden).
The measurements taken on the Faroe Islands and in St.~Petersburg used again the EDR, which was also used to make the recordings in Fig.~\ref{fig:spectrum}\textbf{a}.
In addition, we also include three recordings provided by the Swedish TSO in three undisclosed distinct locations in Sweden for the month of January 2020.
These locations are reported simply as \textit{north Sweden}, \textit{middle Sweden}, and \textit{south Sweden}, as provided by the Swedish TSO.

\begin{figure}[t]
 \includegraphics[width=\linewidth]{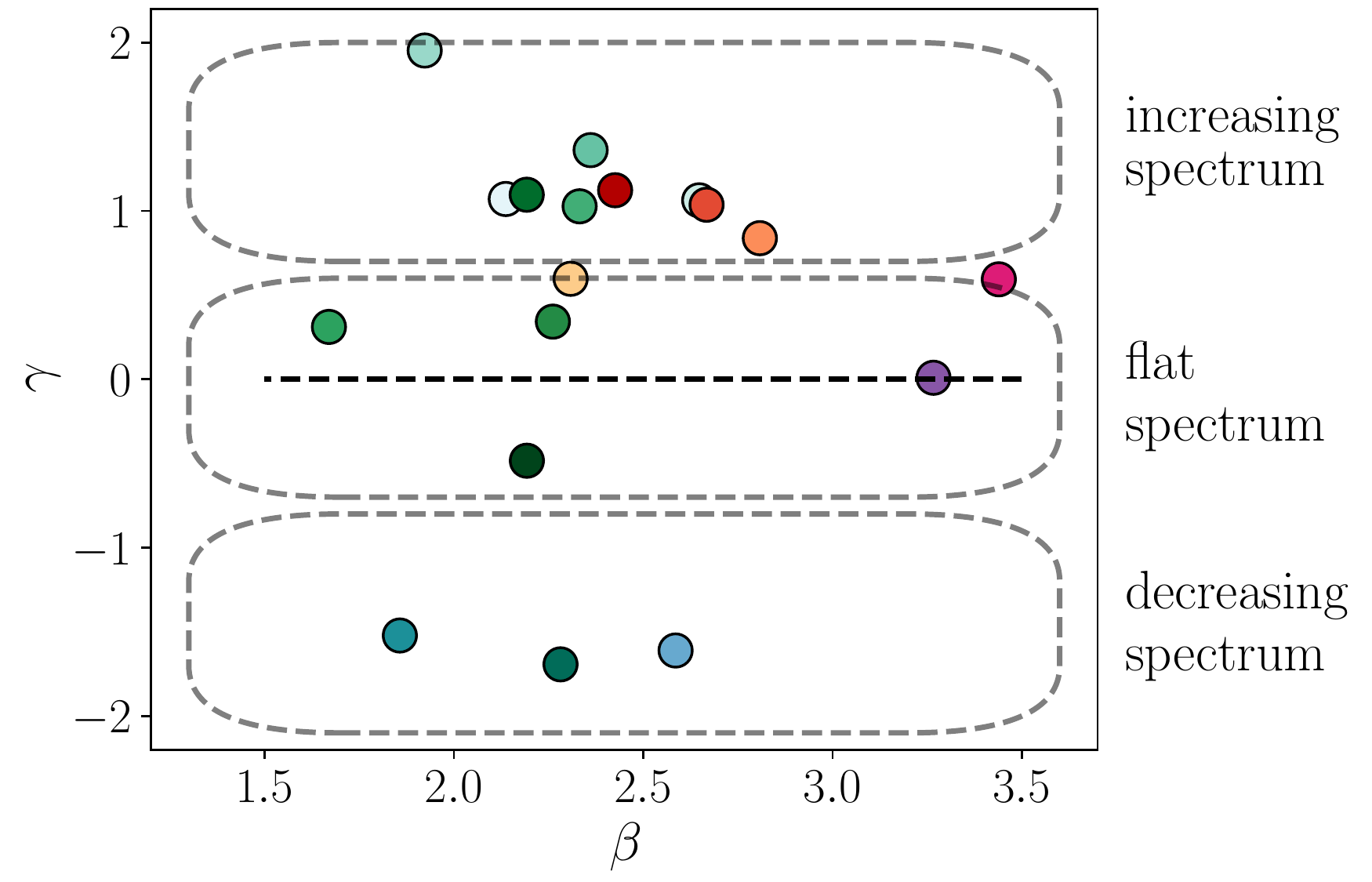}
 \caption{\textbf{Spectral measurements on short time scales are either positively, negatively or not correlated.} We plot estimates of the power law exponents $\beta$ (long time scale) and $\gamma$ (short time scale).}\label{fig:scatter}
\end{figure}

Let us systematically investigate the different slopes $\beta$ and $\gamma$ of the spectral decays observed in Fig.~\ref{fig:spectrum} as a scatter plot in Fig.~\ref{fig:scatter}, see also Supplementary Material for numerical details.
First, we note that all measurements consistently report a decaying power law with exponent $\beta \sim 2.5$.
However, the dynamics at short time scales, given by $\gamma$ is more varied ($\gamma \sim -2...2$): 
We identify three regimes: The Swedish recordings display a continuous decay of the spectrum.
Next, we observe several recordings with small absolute values of $\gamma$, implying the spectrum becomes almost flat.
Finally, many recordings, mostly from the Continental European grid, display negative $\gamma$ values and have an increasing spectrum, pointing towards a complex process emerging at these time scales.

Given this surprising increase of the spectrum on short time scales, let us elaborate on why this is almost certainly not a problem of the measurement system but a genuine effect of the underlying power system.
Subsection~\ref{secsub:ObtainingFrequencyRecordings} deals with the EDR data acquisition and analyzes it towards reliability and trust.
Additionally we make the following observations: 
First, the EDR measurements within one grid are consistent: We clearly observe negative $\gamma$ values in Continental Europe at several independent measurement points, namely in Istanbul, Oldenburg, Karlsruhe and Lisbon.
Second, the negative $\gamma$ values are further confirmed for Continental Europe by measurements from Gridradar (Fig.~\ref{fig:spectrum}\textbf{c}), relying on a completely independent hardware and software solution.
Third, the EDR measurements show $\gamma \approx 0$ in some other synchronous areas, such as in Russia (St.~Petersburg) and on the Faroe Islands.

Interestingly, the time series provided by the Swedish TSO has a positive $\gamma$, while the EDR-recorded time series in Stockholm (recorded during a different time period) yields a low negative value of $\gamma$.
As we do not know the reason for this discrepancy, we list some potential causes:
The measurements by the EDR and the Swedish TSO were taken at different times and could hence measure different states of the power system.
Furthermore, the EDR was connected to a power plug in a hotel, i.e. to the low-voltage distribution grid used for residential areas.
Meanwhile, the Swedish TSO has access to measurements at power plants or directly at the high-voltage transmission grid.
While the frequency is typically a global property, the negative $\gamma$ values could possibly arise from the complex dynamics of components at the local distribution layer, such as inverters and non-linear loads.
Finally, we do not have access to the raw data recorded by the Swedish TSO and hence cannot infer anything about their measurement set-up, filtering etc, while we discuss the limits of our measurements in detail below.

\subsubsection{Multifractality and correlations}
In our sample trajectories in Fig.~\ref{fig:trajectories} we already noted a certain (multi)fractal structure to the power-grid frequency recordings.
One natural question arising is: What is the exact nature of the fractal or multifractal structure that underlies power-grid frequency?

Power-grid systems are complex dynamical systems affected and impacted by various endogenous and exogenous pressures from both within (control mechanics, thermal effects, oscillations) and from outside (changes in generation and consumption, weather, failures).
Every impactful element acts at various different time scales and these time scales range from years (ever-growing generation), to months (weather cycles), to weeks (workweek-weekend cycles), to hours (market and dispatch changes), to minutes (generation adjustments, automated control), to seconds (dynamic interference, travelling waves, occasional failures), to milli-seconds (electrodynamic effects)~\cite{heide2010seasonal,machowski2020power}.

To best describe the power-grid frequency on the very short time scales, we model the time series as a combination of stochastic processes.
To that end, we focus on the incremental structure of our recordings $f(t)$, given by
\begin{equation}\label{eq:increments}
  \Delta f_r = f(t+r) - f(t),
\end{equation}
where $r$ is our incremental lag.
In a stochastic process point-of-view it is usually denoted ``increments''.
These increments carry information relating directly to the process' fastest time scales, i.e. a process changes on a time scale of order $r$, as well as information relating to the existence of memory and fractality in the recordings.

One of the most common characteristics of the incremental time series $\Delta f_r$ is the ``scaling'' of the increments in relation to the incremental lag $r$.
This is commonly encapsulated in the structure function $S_q(\Delta f_r, r)$ 
\begin{equation}
  S_q(\Delta f_r, r) = \langle \Delta f_r^q \rangle \sim r^{qh(q)},
\end{equation}
where $h(q)$ is the generalized Hurst exponent, also known as Hölder exponent, which captures the change in the (non-centered, non-normalized) statistical moments, where $q$ is the power of the statistical moment.
$h(q)$ is related to the scaling exponent $\tau(q)$, which is often used in turbulence analysis, as follows: $\tau(q) = qh(q) - 1$~\cite{ausloos2012generalized}.

We utilize the generalized Hurst exponent for two purposes:
To investigate correlations and to explore (multi)fractality:
For an uncorrelated Brownian motion, we have $h(2)=0.5$, while a correlated process has $h(2)>0.5$ and an anti-correlated process $h(2)<0.5$.
Furthermore, a (mono)fractal process is one where the function $h(q)$ is essentially constant, i.e. independent of $q$, while a multifractal process is characterized by a $q$-dependent generalized Hurst exponent $h(q)$.

If the original time series yields a $q$-dependent $h(q)$, i.e. a multifractal process, this could either be due to different long-range correlations or due to a broad probability distribution.
To ascertain how this multifractality emerges, we shuffle the power-grid frequency recordings, i.e. randomize the position of each time series entry.
We can then calculate the generalized Hurst exponents $h^{\mathrm{shuffled}}(q)$ for the shuffled data.
Since shuffling the data removes correlations, time series that are \textit{not} distributional multifractal will have $h(q) = h(2) = 1/2$, i.e., they will be uncorrelated at all scales.
If, on the contrary, we find that $h(q)$ is non-constant, this is an indication of distributional multifractality.

\begin{figure}[t]
 \includegraphics[width=\linewidth]{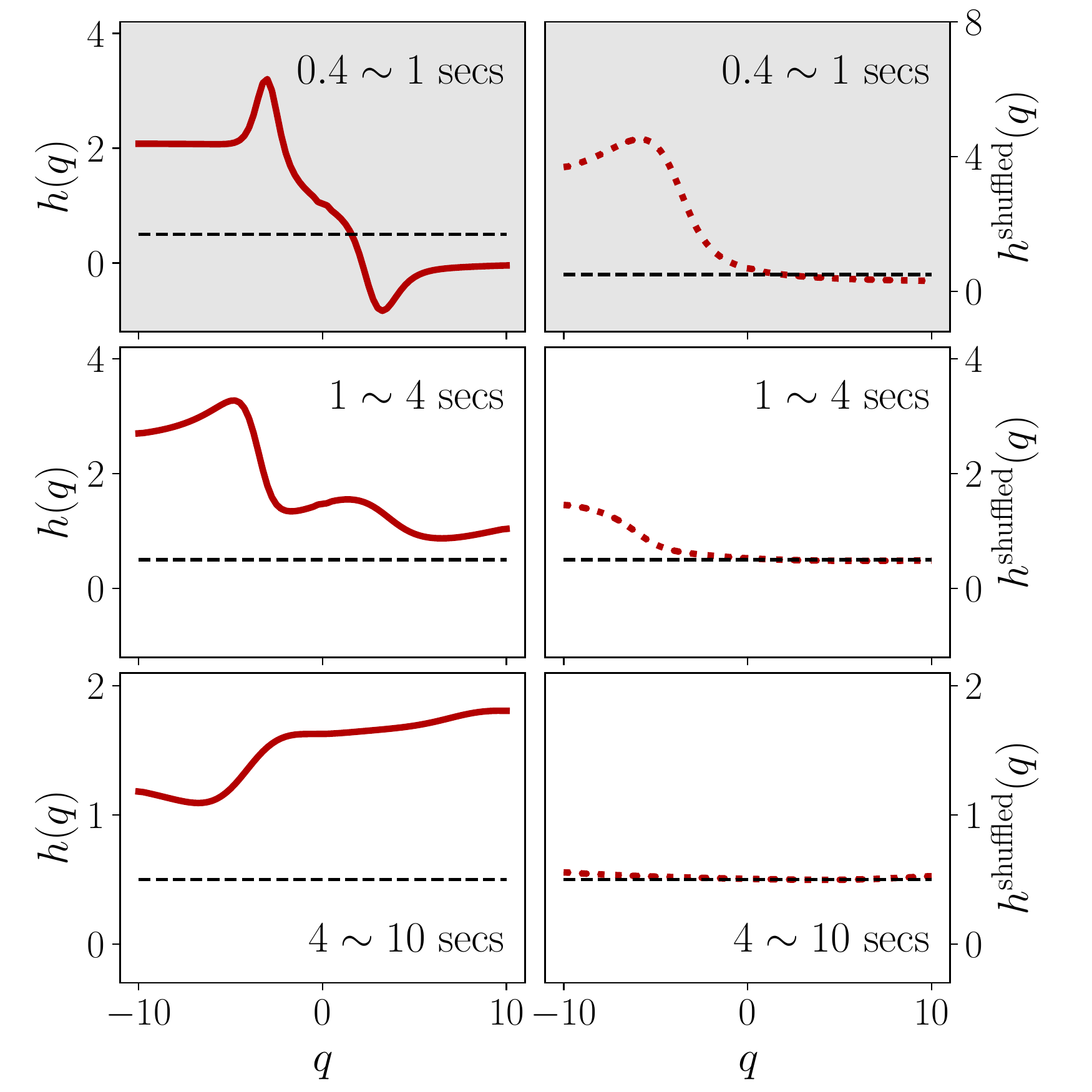}
 \caption{\textbf{Frequency recordings are multi-fractal and anti-correlated on short and correlated on longer time scales.} We plot the generalized Hurst exponent $h(q)$ of power-grid frequency recordings from the Lisbon recording in Continental Europe at different time scales, from top to bottom:
 $0.4\sim1\,$s, $1\sim4\,$s, and $4\sim10\,$s.
 The smallest time scales shows a non-constant exponent $h(q)$, which points to the existence of a large number of degrees of freedom in the system.
 As the time scales become larger, the system gets progressively more ordered, as fluctuation are quenched by the strong synchrony in the system.
 At a time scale of $4\sim10\,$s the shuffled grid frequency recording becomes indistinguishable from noise ($h(q)=0.5$)~\cite{RydinGorjao2022}.}\label{fig:h(q)}
\end{figure}

Computing the generalized Hurst exponent reveals again a complex structure of the time series on the short time scale, see Fig.~\ref{fig:h(q)}: On a time scale of $0.2$ to $1.0$ second, the generalized Hurst exponent is clearly dependent on $q$, indicating a multifractal process, coherent with earlier observations (Fig.~\ref{fig:trajectories}).
This multifractality persists even when shuffling, indicating a distributional multifractality.
For longer time scales, particularly above $4$ seconds, the time series becomes mono-fractal (constant $h(q)$).
Interestingly, the correlation changes over the time scales:
On short time scales of $0.4$ second to $1$ second, the power-grid frequency is anti-correlated as indicated by a small value of $h(2)<0.5$.
For these time scales, essentially, any disturbance on the grid away from the reference frequency of $\SI{50}{\hertz}$ or $\SI{60}{\hertz}$ is followed by an opposite effect.
While we do not expect control actions to play a major role~\cite{machowski2020power} on this time scale, the effect is similar, any disturbance is counteracted, leading to a frequency time series close to the reference value.
A possible explanation for the emerging anti-correlation could be the self-regulating effect of frequency dependent  loads~\cite{hill1982stability}, i.e. a frequency drop would be met by a reduced load and thereby counteract the frequency change.
Meanwhile, the time series becomes positively correlated on longer time scales of several seconds and more, as indicated by a large value of $h(2)>0.5$.
This persistent behavior aligns with our expectations of a power system subject to load and generation ramps on time scales of seconds to minutes and an overall inertial system.

Let us briefly discuss the role of distributional multifractality before then utilizing the insights about correlations in the time series to reproduce the observed spectrum.
We do not know for certain why and how the multifractality arises in the frequency recordings discussed.
However, we do know that one mechanism giving rise to a multifractal time series is the superposition of several processes at a shorter time scale, formalized under the concept of superstatistics by Beck and Cohen~\cite{beck2003superstatistics}.
Assuming the observed frequency time series arises from such a superstatistics, it would offer further directions of research.
In superstatistics, a system changes from one state to another while we treat all observations as if they originated from the same state.
In terms of the energy system, these different states of the system could be related to different generation mixes and demand patterns or due to drivers such as superstatistical behavior in wind~\cite{Friedrich2022}.
Note that superstatistics has been discussed on power-grid frequency recordings before~\cite{schafer2018non} and there are further hints that the dynamics of power grid recordings gets more complex on shorter time scales.
In fact, if we consider the temporal scales wherein we observe the strongest evidence of multifractality, i.e. for $t<4$\,s, this agrees well with previous results~\cite{RydinGorjao2021a,RydinGorjao2021b}, wherein power-grid frequency increments are shown to be well described by a convolution of Gaussian statistics with a parametric function of varying variance.
All of this points to even more research being necessary to understand the complexity of power-grid frequency time series on this short time scale.

\subsubsection{Stochastic process}\label{para:StochasticProcess}

\begin{figure}[t]
 \includegraphics[width=\linewidth]{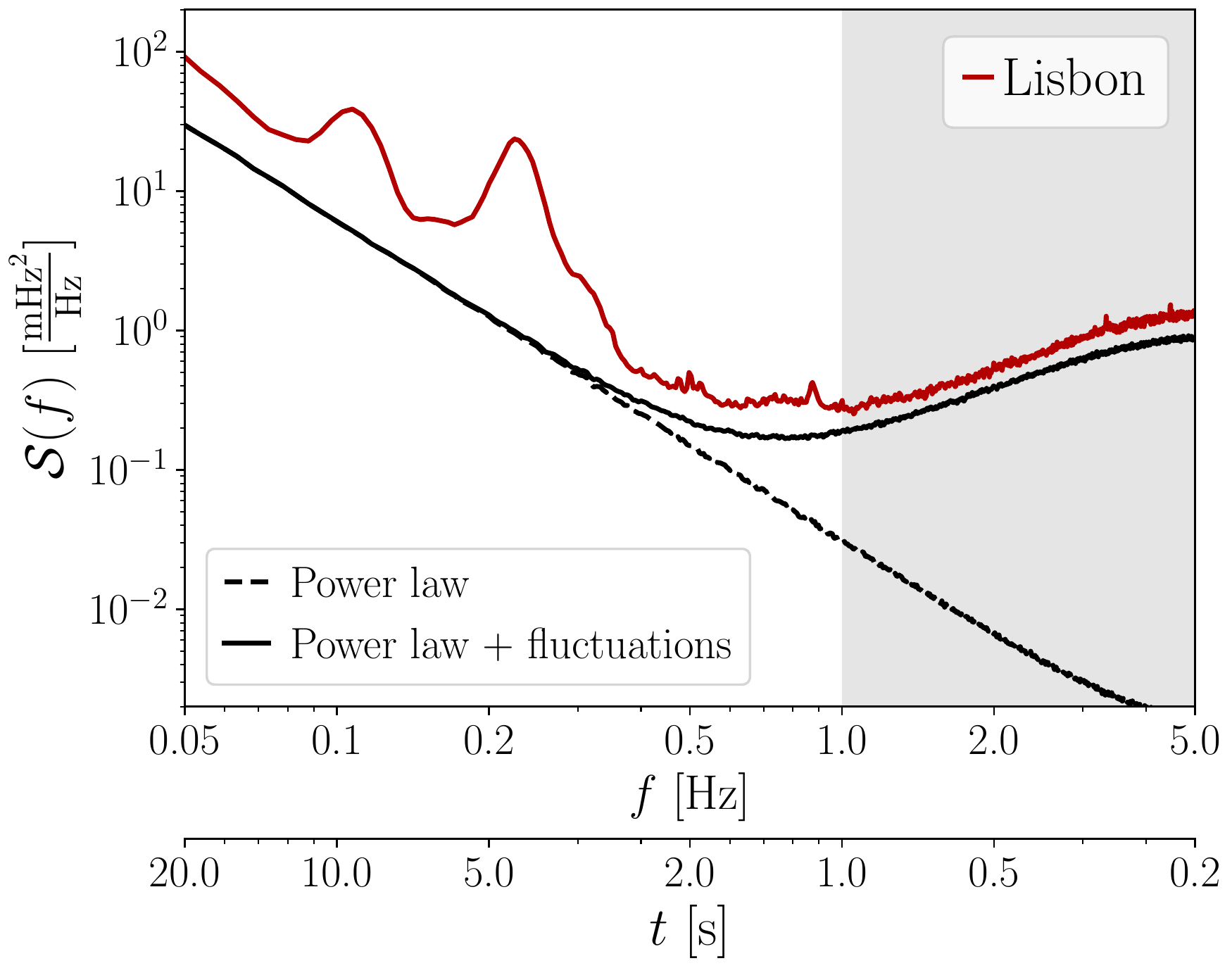}
 \caption{\textbf{Reproducing the spectrum using an Ornstein--Uhlenbeck process with added correlated short-term fluctuations.}}\label{fig:spectrum_synthetic}
\end{figure}

Let us return to the observation of an increasing spectrum and try to explain this via physical processes on the short time scale.
We have established from our generalized Hurst exponent analysis that the frequency is positively correlated on a longer time scale of several seconds and negatively correlated on shorter time scales.
For now we consider simple mono-fractal processes and explore how these can reproduce the observed spectrum.
Indeed, a very simple stochastic process can already reproduce such a spectrum, namely combining a fractional, positively correlated Ornstein--Uhlenbeck process with added, negatively correlated fluctuations, see Fig.~\ref{fig:spectrum_synthetic}.
Before discussing Fig.~\ref{fig:spectrum_synthetic} in detail, let us motivate the usage of an Ornstein--Uhlenbeck process:

Following earlier work~\cite{schafer2018non,gorjao2020data}, we may model the frequency dynamics of the lossless high-voltage grid on the short time scale as
\begin{equation}
  \begin{split}
  \frac{\text{d}}{\text{d}t}\theta_{i}&=\omega_{i},\\
  M_{i}\frac{\text{d}}{\text{d}t}\omega_{i}&=P_{i}+\sigma_{i}\xi_{i}-g_{i}\omega_{i}+\sum_{j=1}^{N}K_{ij}\sin\left(\theta_{j}-\theta_{i}\right),
  \end{split}
\end{equation}
where we assume that each node $i$ in the grid behaves approximately like a synchronous machine whose state is characterized by its voltage phase angle $\theta_{i}$ and angular velocity $\omega_{i}=2\pi\left(f_{i}-f_{R}\right)$, which is proportional to the deviation of the frequency from the reference frequency of $f_{R}=\SI{50}{\hertz}$ or $f_{R}=\SI{60}{\hertz}$.
The dynamics of the network is then determined by the following parameters: inertia $M_{i}$, mechanical power $P_{i}$ (positive for effective generators and negative for effective consumers), fluctuations $\xi_{i}$ with amplitude $\sigma_{i}$, damping $g_{i}$ and the coupling matrix $K_{ij}$.

As we typically do not have the precise information about the full network topology, all nodal parameters and states, we want to simplify this to an equation for the bulk angular velocity $\bar{\omega}:=\frac{\sum_{i=1}^N M_i \omega_i}{\sum_{i=1}^N M_i}$ by assuming a homogeneous damping to inertia ratio, $c=g_{i}/M_{i}$~\cite{weixelbraun2009dynamic} and balanced average power $\sum_{i=1}^{N}P_{i}=0$.
We obtain the aggregated swing equation as
\begin{equation}
  \frac{\text{d}}{\text{d}t}\bar{\omega}=-c\bar{\omega}+\bar{\sigma}\bar{\xi}\left(t\right), \label{eq:bulk_dynamics}
\end{equation}
with aggregated fluctuation amplitude $\bar{\sigma}$ and aggregated fluctuations $\bar{\xi}$, which depend on the precise nature of the noise~\cite{schafer2018non}.
The bulk frequency dynamics given in Eq.~\eqref{eq:bulk_dynamics} describes the average behavior in one synchronous area, such as the Continental European grid, neglecting specific local properties.
Crucially, we have simplified the non-linear network dynamics to a relatively simple Ornstein--Uhlenbeck process.
A main difference to a standard Ornstein--Uhlenbeck process is that the aggregated fluctuations $\bar{\xi}$ are typically not well described by ordinary Brownian motion but instead the effective fluctuations acting on the system are either heavy-tailed itself or arise from a superposition of several stochastic processes, so-called superstatistics~\cite{beck2003superstatistics,schafer2018non}, mentioned earlier.

Regardless of the precise nature of the fluctuations $\bar{\xi}$, for an Ornstein--Uhlenbeck process, we would expect a spectrum following a simple power-law (dotted line in Fig.~\ref{fig:spectrum_synthetic})
In order to obtain an explicit expression for the power-spectral density, recall that the energy of a time series is given by
\begin{equation}\label{eq:energy}
  E := \int\limits_{-\infty}^{\infty}\left|x(t)\right|^2 \mathrm{d}t.
\end{equation}
Parseval's theorem dictates that
\begin{equation}
  \int\limits_{-\infty}^{\infty}\left|x(t)\right|^2 \mathrm{d}t = \int\limits_{-\infty}^{\infty}\left|\hat{x}(f)\right|^2 \mathrm{d}f,
\end{equation}
with $\hat{x}(f)$ the Fourier transform of $x(t)$, i.e.,
\begin{equation}
  \hat{x}(f) = \int\limits_{-\infty}^{\infty}x(t)e^{-i2\pi ft} \mathrm{d}t,
\end{equation}
with $i$ the imaginary unit.
Using Parseval's theorem, we see that the integrand of the energy in Eq.~\eqref{eq:energy} is simply the spectral density $\mathcal{S}(f)$ given as
\begin{equation}\label{eq:spectral_density_def}
  \mathcal{S}(f) := \left|\hat{x}(f)\right|^2.
\end{equation}
In the following, we are interested in a solution of the aggregated swing equation~\eqref{eq:bulk_dynamics} in Fourier space.
Defining $\hat{\bar{\omega}}$ and $\hat{\bar{\xi}}$ as the Fourier transforms of the bulk angular velocity $\bar{\omega}$ and the aggregated noise amplitude $\bar{\xi}$ respectively and noting that $c$ and $\bar{\sigma}$ are constants, yields 
\begin{equation}
  if\hat{\bar{\omega}} + c\hat{\bar{\omega}} = \bar{\sigma}\hat{\bar{\xi}},
\end{equation}
which we can express in relation to $\hat{\bar{\omega}}$ as
\begin{equation}
  \hat{\bar{\omega}} = \frac{\bar{\sigma}\hat{\bar{\xi}}}{if + c}.
\end{equation}
Inserting this into our expression of the power spectral density in Eq.~\eqref{eq:spectral_density_def} results in the power spectral density of $\bar{\omega}$ as
\begin{equation}
  \mathcal{S}(f) = \frac{\bar{\sigma}^2|\hat{\bar{\xi}}|^2}{f^2 + c^2}.
\end{equation}
Since $\bar{\sigma}^2$ and $c^2$ are constants, the spectral density dependence on the frequency $f$ is solely determined by the $1/f^2$-term and the noise contribution $\bar{\xi}$.
If the noise is homogeneous in the frequency, i.e. it is white noise, then we obtain a power-law $\mathcal{S}(f)\sim f^{-2}$, i.e. $\mathcal{S}(f)$ decays with an exponent of $-2$, consistent with our observations.

To reproduce the observed spectrum on short time scales, we consider the addition of negatively correlated Brownian fluctuations $\eta(t)$ so that the full dynamics reads
\begin{equation}
  \bar{\omega}(t) = -c\int\limits_{0}^{t}\bar{\omega}(t')\mathrm{d}t' + \sigma\int\limits_{0}^{t}\bar{\xi}(t')\mathrm{d}t'
   + \eta(t)
\end{equation}
Indeed, the experimental spectra can be reproduced by this synthetic process, see Fig.~\ref{fig:spectrum_synthetic}.
We observe that instead of a continued decaying power law, both the synthetic process but also the measurements (in this case from Lisbon) do change qualitatively at the one second time scale with a positive slope of the spectrum.

In this section, we have demonstrated new characteristics of the power-grid frequency dynamics on the time scale below one second.
Such analysis relied on the availability of high-resolution frequency readings.
In the next section, we discuss how such recordings were obtained and validated.

\subsection{Obtaining high-quality recordings from the power grid}\label{secsub:ObtainingFrequencyRecordings}
To explain and investigate the origin of our frequency measurements, we begin by defining the power-grid frequency.
We model the power grid voltage curve $u(t)$ at a measurement location in the grid by
\begin{equation}
  u(t) = A \sin(\theta(t)),
\end{equation}
wherein $\theta(t)$ is the instantaneous phase at time $t$ and $A$ the amplitude.
The instantaneous frequency of this signal is then defined as
\begin{equation}\label{eq:instFreq}
  f_\mathrm{inst}(t) = \frac{1}{2\pi} \frac{\mathrm{d} \theta(t)}{\mathrm{d}t}.
\end{equation}

With $f_\mathrm{inst}(t)$ we have the definition of the power-grid frequency.
Unfortunately though, $f_\mathrm{inst}(t)$ is not directly observable, so we need a measurement process to determine some repeated approximations of $f_\mathrm{inst}(t)$ from $u(t)$.

In the following, we will layout this process in the subsections~\ref{sec:DataAcq}, \ref{sec:CalcFreq}, and \ref{sec:FreqEst} making use of already published developments and we focus on accuracy in \ref{sec:AccuracyFreq}.
We extend the works by addressing the impact of signal disturbances on $\mathcal{S}(f)$ in \ref{sec:InfluenceOfDist}.

\subsubsection{Data acquisition}\label{sec:DataAcq}
Our measurement device, the Electrical Data Recorder (EDR), has been introduced some years ago and is being developed continuously, see also~\cite{maass2013first, maas2015data, jumar2021database, foerstner2022experimental}.
In the following, we provide a general overview over the data acquisition and frequency calculation.
We then specify how to derive frequency data with high temporal resolution.

\begin{figure}[t]
 \includegraphics[width=\linewidth]{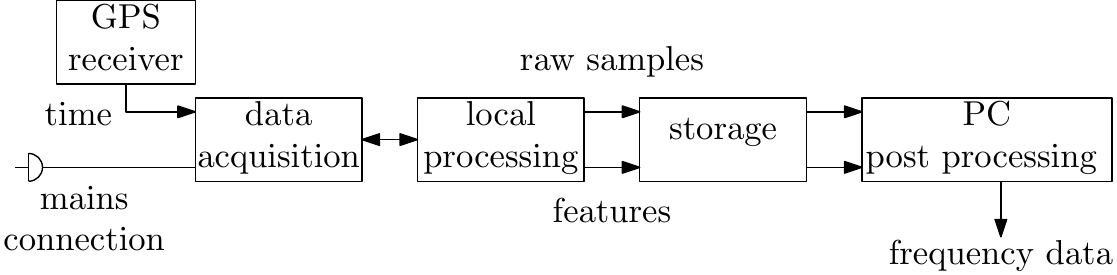}  
 \caption{\textbf{Conceptual illustration of the data acquisition and processing chain.}
 Data is acquired, processed and stored on the Electrical Data Recorder device.
 Both, raw samples and processed data are available for in-depth analysis on a PC.
 The EDR also supports remote data storage, live monitoring, current measurements etc. which are not depicted here.
 See~\cite{EDRprep2014,EDR_Rework2019,maass2013first} for details.}\label{fig:edrProcChain}
\end{figure}

The EDR captures the voltage $u(t)$ at a given connection point with a sampling frequency of \SI{25}{kHz}.
Using the time-information provided by an GPS satellite receiver, captured data is assigned a standardized time and date information (UTC time stamp).
Via the time stamps, the acquisition and the subsequent data processing inherit the precision of GPS as a local standard of the second.
This allows to directly overlay and compare data from different locations without the need for manual alignment in post-processing.
Based on the recording of the $u(t)$ time series, we extract features, such as the signal frequency.
We visualize the data processing chain in Fig.~\ref{fig:edrProcChain}.

\subsubsection{Calculation of the frequency}\label{sec:CalcFreq}
Signal frequency estimation in the EDR begins by passing the digitized waveform $u(t)$ through a linear phase low-pass Finite Impulse Response (FIR) filter.
For \SI{25}{\kilo\hertz} sampling rate and \SI{50}{\hertz} nominal frequency the filter has a length of $L=200$ and a \SI{3}{\decibel} cut off frequency close to \SI{50}{\hertz}.
Subsequently, the EDR determines period-wise frequency readings $f_\mathrm{p}(k)$ by measuring the time between zero-crossings:
\begin{equation}
  f_\mathrm{p}(k) = \frac{1}{t_\mathrm{zc}(k)-t_\mathrm{zc}(k-1)}
\end{equation}
The series of zero-crossing times $t_\mathrm{zc}(k)$ are designated by instances where $\theta = 0, 2\pi,\dots$, and $k$ is the index of a specific zero-crossing.
Looking at zero-crossings makes the sampling times $t_\mathrm{zc}$ dependent on the instantaneous phase of $u(t)$.
For any practical signal, the elements of $t_\mathrm{zc}(k)$ are not equidistantly spaced on the time axis.

\begin{figure}[t]
  \includegraphics[width=\linewidth]{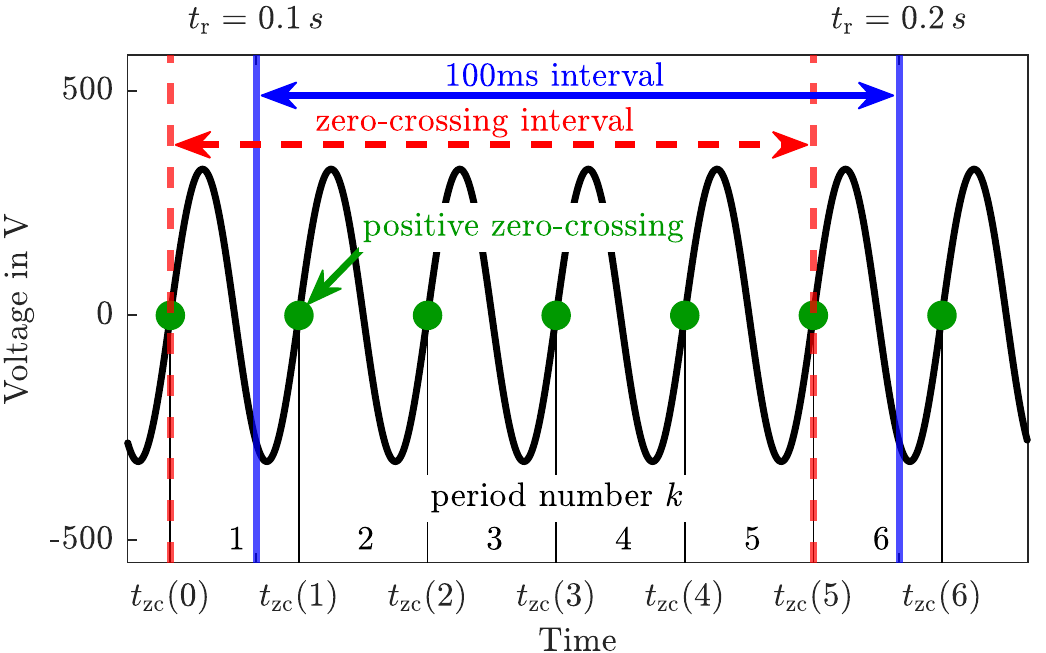}
  \caption{\textbf{Zero-crossing method}.
  Illustration of the zero-crossing positions relative to a \SI{100}{\milli\second} time interval.}
  \label{fig:ZeroCrossing}
\end{figure}

Since data with nonuniform sampling is subsequently harder to analyze and impossible to compare directly between locations and with other data sources, the acquired information needs to be mapped to regular time intervals.
We call this desired regular output ``frequency reports''~$f(n)$, where $n$ points to a specific element of the series.
Each report carries a time stamp $t_\mathrm{rep}$ that is derived from a GPS-synchronized clock.
The time between $t_\mathrm{rep}(n)$ and $t_\mathrm{rep}(n+1)$ is called ``report interval''.
In the following, we describe how we calculate $f(n)$ from $f_\mathrm{p}(k)$.

\subsubsection{Frequency estimation -- increasing time resolution}
\label{sec:FreqEst}
As shown in the previous section, frequency estimates are especially interesting at time scales around and below one second, as they reveal new characteristics.
Obtaining theses estimates requires special attention with regard to temporal alignment.

For report intervals of one second and larger it seems plausible to directly aggregate $f_\mathrm{p}$ and assign the results to $t_\mathrm{r}$ - ignoring the slight misalignment due to the position of the zero-crossings.
We call this approach the full-period-method (fp).
Fig.~\ref{fig:ZeroCrossing} illustrates the aggregation window used by this approach in red and the report interval in blue.
The frequency estimation $f_\mathrm{fp}$ for the aggregation interval $n$ is found by averaging over the $K$ periods of the aggregation interval
\begin{equation}
  f_\mathrm{fp}(n) = \frac{1}{K(n)} \sum_{k=1}^{K(n)}{f_\mathrm{p}(k,n)}.
\end{equation}
Due to the time misalignment, this kind of estimation causes an error that depends on both the phase $\theta(t)$ at the interval borders and the derivative of the instantaneous frequency $f_\mathrm{inst}$ within the interval.
The error thereby scales linearly with the latter.
In the worst case, a full period in the average does not actually belong to the report interval $n$.
For one second reports this equals \SI{2}{\percent} of the involved data and is usually tolerated.
For \SI{100}{\milli\second} reports this figure increases to \SI{20}{\percent} and causes intolerable deviations during transient frequency events.

To address this challenge we introduced the partial period method (pp) in~\cite{foerstner2022experimental} that performs a local re-sampling of the data to the regular report intervals.
Contributions of the border-periods (i.e. periods 1 and 6 in Fig.~\ref{fig:ZeroCrossing}) are partially included in the average according to their temporal overlap with the report interval marked in blue:
\begin{equation}
  f_{\mathrm{pp}}(n) = \frac{1}{K(n)-2 + w_1 + w_K} \sum_{k=1}^{K(n)} {w_k f_\mathrm{p}(k,n)},
\end{equation}
where $K$ is the number of periods that are in touch with the aggregation interval $n$ and $w_k$ are weights for the individual periods.

The first weight is given by
\begin{equation}
  w_1 = (t_\mathrm{zc}(1) - t_\mathrm{rep}(n-1))\cdot f_\mathrm{p}(1)
\end{equation}
and the last weight is given by
\begin{equation}
  w_K = (t_\mathrm{rep}(n) - t_\mathrm{zc}(K-1))\cdot f_\mathrm{p}(K),
\end{equation}
while all periods in between are equally weighted with $w_{k=2\dots K-1}=1$.

While the pp-method eliminates the majority of shift related estimation errors, subtle deviations remain.
The main reason is that the approach assumes that $f_{\mathrm{inst}}$ is constant during a period.
This condition is hardly satisfied for any real signal.
Nevertheless, the estimation error in dynamic conditions is drastically improved compared to the fp-method~\cite{foerstner2022experimental}.
The partial period method therefore practically solves the temporal alignment problem.

\subsubsection{Accuracy of frequency measurements}\label{sec:AccuracyFreq}
Since the individual period measurements are only as good as the underlying time reference, we use the GPS receivers' Pulse Per Second (PPS) signal as a local frequency standard.
The error in frequency $\Delta f$ relates to the uncertainty in the time $\Delta t$ via
\begin{equation}
  \Delta f = \left|-(T=1/f_\text{nom})^{-2}\right|\Delta t.
\end{equation}
The edge-to-edge error of our receiver (Garmin GPS-18LVC) and PPS acquisition is less than \SI{50}{\nano\second}, which translates to a worst-case frequency error of \SI{125}{\micro\hertz} in the \SI{50}{\hertz} case.
Experimentally, we further verify that the frequency estimation error in steady state conditions over a 10 minute interval is only \SI{26.7}{\micro\hertz} at \SI{50}{\hertz} within a \SI{95.4}{\percent} confidence interval~\cite{EDRprep2014}.
In dynamic conditions however, the estimation errors are dominated by the principles of the zero-crossing method.
A detailed analysis of the estimation errors in time domain can be found in~\cite{foerstner2022experimental}.
For the frequency ramps defined in the standard for Phasor Measurement Units (IEEE 60255-118-1) we found the error limit of our method to be $\pm\SI{0.6}{mHz}$.

\subsubsection{Influence of signal disturbances}\label{sec:InfluenceOfDist}
As shown in \ref{para:StochasticProcess} and Fig.~\ref{fig:spectrum_synthetic}, the observed PSDs can be reproduced using a stochastic process, composed of an Ornstein--Uhlenbeck process (including white noise) and added correlated fluctuations, where the latter lead to the increasing spectrum on a short time scale.
We thus investigate how our processing chain deals with disturbances in the signal.
By doing so, we also ensure that no artificial signal is introduced in the system.

\begin{figure}[t]
  \includegraphics[width=0.95\linewidth]{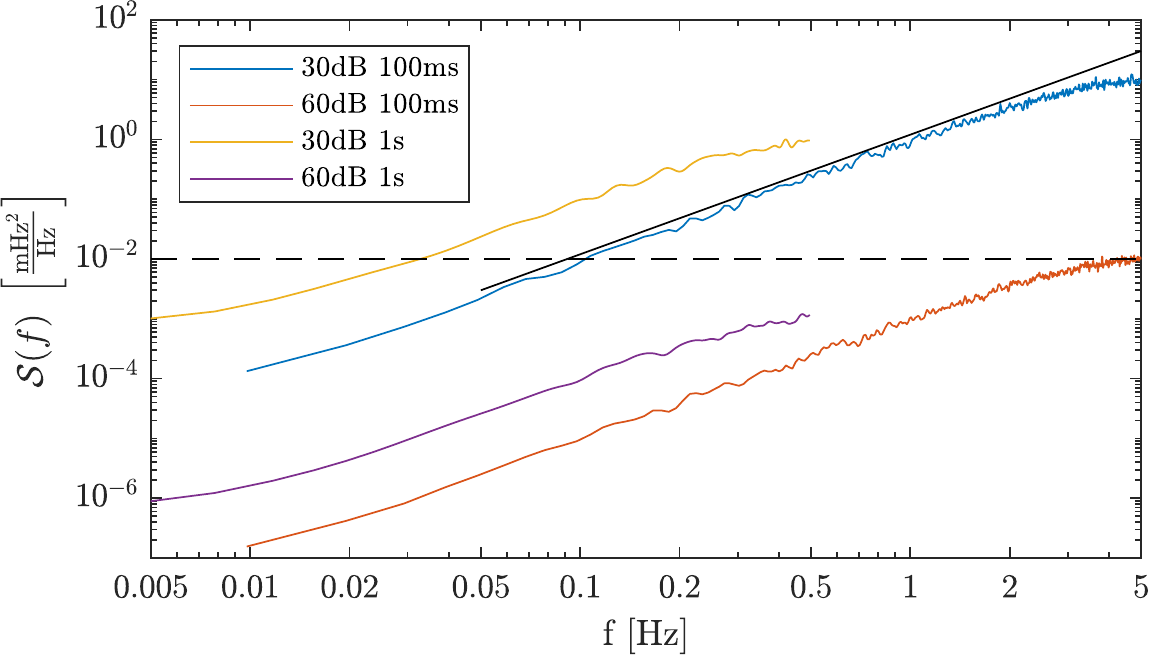}
  \caption{\textbf{Power spectral density of $f$ for the partial period zero-crossing method}.
  Input is a sine wave with additional white noise of different levels.
  PSD estimates saturate for long period duration and rise nearly quadratic (solid black line) towards shorter periods.
  For 60~dB SNR the dashed line indicates the maximum level of $\mathcal{S}(f)$ at \SI{0.01}{\milli\hertz\squared\per\hertz}.}
  \label{fig:noiseTest}
\end{figure}

From reviewing the signal chain, disturbances superimposed on $u(t)$ are the most likely source of the observed fluctuations, since adding a random signal $e_\mathrm{u}(t)$ to $u(t)$ will impact how well the period boundaries can be determined.
The relation between such voltage noise $e_\mathrm{u}$ and noise in the period duration (time) measurement $e_\mathrm{t}$ is given by the gradient of $u(t)$ at the zero-crossing:
\begin{equation}
  e_\mathrm{t}(k) = e_\mathrm{u} \frac{d}{dt}u(t = t_\mathrm{ZC}(k)) = 2\pi A f_0 e_\mathrm{u}(k).
\end{equation}
Voltage noise therefore translates linearly to noise in period duration (phase noise), scaled by power-grid frequency and the signal amplitude.
However, if $e_\mathrm{u}$ is not random, the influence might vanish.
This is especially the case for harmonics, where $e_\mathrm{u}$ is periodic in multiples of~$f$.

To gain an impression on the influence of noise, we conduct the following computer experiment.
We generate a sinusoidal signal, superimpose it with white Gaussian noise and process it with the estimation algorithm to obtain a series of frequency readings $f(n)$.
We then determine the Welch PSD estimates $\mathcal{S}(f)$ of this series as done for the measured frequency time series.
Results are shown in Fig.~\ref{fig:noiseTest} from which we make three major observations: 

\begin{figure}[t]
 \includegraphics[width=\linewidth]{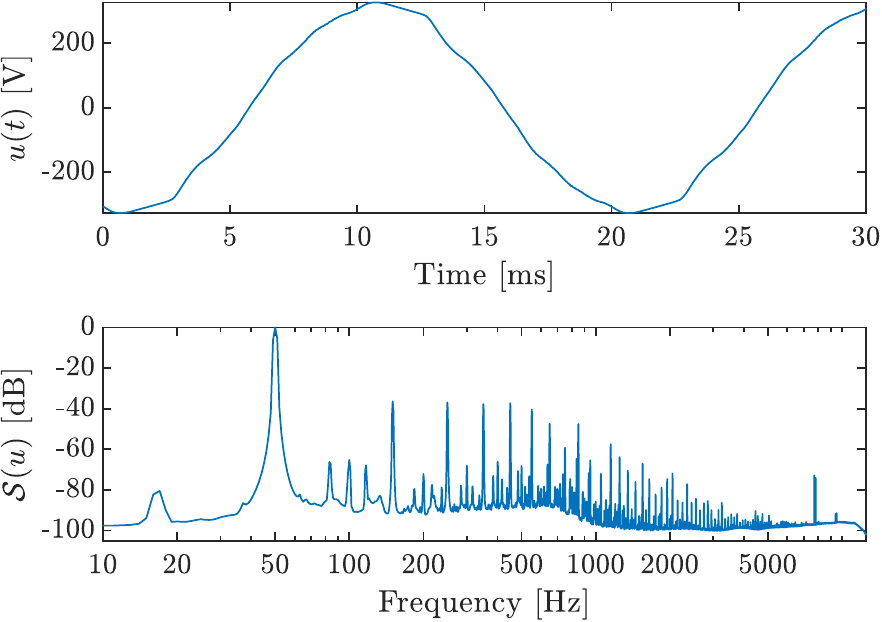}  
 \caption{\textbf{Captured voltage waveform in Stockholm}.
 Top: Example of the captured voltage waveform $u(t)$.
 Data: Stockholm 9\textsuperscript{th} of May 2019, 20:35.
 Bottom: Power spectrum of one hour of data (Welch estimate, resolution \SI{1}{Hz}, Hann-Window).
 Signal to Noise Ratio (SNR) including distortion: \SI{29.6}{\decibel}.
 This is worst case SNR observed for EDR measurements.}\label{fig:CombinedWaveformAndVoltageSpectrumPlot}
\end{figure}

1) The curves for \SI{30}{\decibel} and \SI{60}{\decibel} signal-to-noise-ratio (SNR) are similar in shape but vertically shifted by three orders of magnitude.
The shift corresponds exactly to the \SI{30}{\decibel} difference in noise excitation levels.
Therefore, the noise levels in $f$ and $u(t)$ can be directly related.

2) $S(f)$ rises approximately proportional to quadratic with $t^{-2}$ which is expected from frequency demodulator theory~\cite{Proakis2013}.
Since the ZC-algorithm averages $K$ period measurements, we would assume that the total noise amplitude of $f$ scales with $1/\sqrt{K}$.
However, due to the increase in noise density, the total noise amplitude in $f$ increases linearly with $K$.
Changing the resolution from \SI{1}{\second} to \SI{100}{\milli\second} translates to a tenfold increase in noise amplitude ($\sigma=\SI{16}{\micro\hertz} ~\mathrm{vs.}~ \SI{160}{\micro\hertz}$ in the $\SI{60}{\decibel}$ case, and $\sigma=\SI{0.48}{\milli\hertz} ~\mathrm{vs.}~ \SI{4.9}{\milli\hertz}$ in the $\SI{30}{\decibel}$ case).
Hence, the data with higher temporal resolution contains more noise in absolute terms.

3) For smaller values of $f$ we observe a slight flattening of $\mathcal{S}$ indicating a possible convergence to a constant value.
The shallow part indicates some feed-through of white phase noise, which is even more pronounced for the full-period-method (shown in Supplementary Material).
We find the source of this effect in the violation of the previously introduced assumption:
The instantaneous frequency is not constant during each period.
Compared to the noise contributions from the high-frequency-end of the spectrum, the feed-trough has minuscule effect on the results.

\subsubsection{Quantification of disturbance influence}
So far, we have characterized the effects of noise on the frequency estimation.
We still need to quantify what noise contribution is to be expected in an actual measurement environment so that we can establish a boundary of trust for our results.
For this we need to a) analyze the distortions in the real world setting and b) characterize the susceptibility of algorithm.

To approach a) we investigate the properties of $u(t)$.
In Fig.~\ref{fig:CombinedWaveformAndVoltageSpectrumPlot} we plot an excerpt of the waveform and the spectrum of the measured voltage.
That majority of the disturbance is in the non-sinusoidal shape of the waveform and therefore in the harmonic content.

From the spectrum of $u(t)$ we can determine the power contained in the disturbances to obtain a Signal-to-Noise (SNR) ratio for the measurements (details in Supplementary Material).
Throughout the campaigns we find a minimum SNR of $\approx\SI{30}{\decibel}$.
However, if we selectively suppress the contribution of the first ten harmonics, the minimum SNR rises to $\approx\SI{60}{\decibel}$.

Additionally to the above, short term transients, caused by local switch operations, will occasionally occur in $u(t)$.
From the time-domain analysis of the algorithm in~\cite{foerstner2022experimental} we know, that such transients will cause large (usually \SI{0.1}{\hertz} and above) single point frequency deviations.
Since a pulse in Dirac sense translates to a constant in frequency domain, those transients might cause slightly elevated $\mathcal{S}(f)$ levels.
However, most of these single point deviations are removed from the data via an outlier filter prior to our investigation.

For b), we already know from the processing chain, that components above \SI{50}{\hertz} are rejected by the initial low-pass filtering.
It is further known that static waveform deformations have no impact on the estimation~\cite{foerstner2022experimental}.
We are therefore convinced that omitting the major harmonic contributions from the SNR estimation is justified (further discussion in Supplementary Material).

Consequently, the expected contribution of ZC-algorithm to the observed PSD estimates follows the \SI{60}{\decibel} curve in Fig.~\ref{fig:noiseTest}.
This means, that in all cases except Oldenburg, the algorithms contribution explains less than one tenth to the observed $\mathcal{S}(f)$.
In the case of Oldenburg the factor is $\approx1/4$, while in most cases it is in the range of $\approx1/100$.
Hence, we are confident that the observed increases of $\mathcal{S}(f)$ are not a measurement artifact but a signal property.

\section{Discussion}
Overall, we uncovered complex behavior in power-grid frequency recordings from high-quality measurements.
Analyzing the data, we made three key observations: 
First, we have seen that the PSD decays approximately following a power law until a time scale of 1 second, thereby displaying a typical complex system property.
Second, we noted distinct dynamics on the shorter time scale with an increasing spectral amplitude, multifractal behavior and a change from positive to negative correlations.
Third, we showed how key properties of the spectrum may be reproduced by describing the power grid as a complex system following an Ornstein--Uhlenbeck process with added fluctuations, motivated mostly from basic principles.
Finally, we described the acquisition processing chain used to acquire the frequency data.
The challenges in terms of temporal alignment and measurement noise arising when moving to high temporal resolution were outlined and we discussed how the EDR-derived data deals with these challenges.

What does this mean for our understanding of complex systems and power system operation and modeling?
To enable any of our observations it was critical to have access to high-quality, highly resolved data.
This stresses the need to monitor power systems closely and also to make these data, including simultaneous and independent measurements in multiple synchronous areas, available for broad analysis and comparison.

Furthermore, we have seen that fluctuations, especially on short time scales, are complex and not simply described as Gaussian white noise but instead seem to follow an anti-correlated and multifractal process.
Possible explanations for the emerging anti-correlation range from effects in the distribution grid, the self-regulating effect of frequency dependent loads~\cite{hill1982stability}, layered network dynamics \cite{tyloo2022layered} to superstatistics.
Regardless of the cause, this complexity should be implemented explicitly by any modeler when simulating and analyzing power systems, especially as an increasing share of renewables and reductions of the total inertia will likely amplify non-standard statistics only further.
Only when we fully model and understand fluctuations, can we design efficient and effective control strategies to ensure stability.

Further research is necessary to fully explain what particular elements in power systems give rise to multifractal phenomena and investigate its potential connection to superstatistics.
On a similar note, complex spectra are observed in other systems~\cite{sim2022offshore} so that we suspect the existence of more non-standard spectra in other time series.

\subsection*{Data availability statement}
The power grid frequency provided by gridradar are not publicly available but are typically made available for scientist upon reasonable request. The remaining frequency data used in this study are openly available at \url{https://power-grid-frequency.org/}. Most code to reproduce the results shown here is made available at github \cite{github}.

\subsection*{Acknowledgments}
We thank Tobias Veith from Gridradar for providing us with data.
This work was supported by the Helmholtz Association under the program ``Energy System Design'' and via grant no. VH-NG-1727 and the program ``Uncertainty Quantification -- From Data to Reliable Knowledge (UQ)'' via grant no.~ZT-I-0029, as well as the Scientific Research Projects Coordination Unit of Istanbul University, Project no. 32990.

\bibliography{bib}

\begin{thebibliography}{49}%
\makeatletter
\providecommand \@ifxundefined [1]{%
 \@ifx{#1\undefined}
}%
\providecommand \@ifnum [1]{%
 \ifnum #1\expandafter \@firstoftwo
 \else \expandafter \@secondoftwo
 \fi
}%
\providecommand \@ifx [1]{%
 \ifx #1\expandafter \@firstoftwo
 \else \expandafter \@secondoftwo
 \fi
}%
\providecommand \natexlab [1]{#1}%
\providecommand \enquote  [1]{``#1''}%
\providecommand \bibnamefont  [1]{#1}%
\providecommand \bibfnamefont [1]{#1}%
\providecommand \citenamefont [1]{#1}%
\providecommand \href@noop [0]{\@secondoftwo}%
\providecommand \href [0]{\begingroup \@sanitize@url \@href}%
\providecommand \@href[1]{\@@startlink{#1}\@@href}%
\providecommand \@@href[1]{\endgroup#1\@@endlink}%
\providecommand \@sanitize@url [0]{\catcode `\\12\catcode `\$12\catcode
  `\&12\catcode `\#12\catcode `\^12\catcode `\_12\catcode `\%12\relax}%
\providecommand \@@startlink[1]{}%
\providecommand \@@endlink[0]{}%
\providecommand \url  [0]{\begingroup\@sanitize@url \@url }%
\providecommand \@url [1]{\endgroup\@href {#1}{\urlprefix }}%
\providecommand \urlprefix  [0]{URL }%
\providecommand \Eprint [0]{\href }%
\providecommand \doibase [0]{https://doi.org/}%
\providecommand \selectlanguage [0]{\@gobble}%
\providecommand \bibinfo  [0]{\@secondoftwo}%
\providecommand \bibfield  [0]{\@secondoftwo}%
\providecommand \translation [1]{[#1]}%
\providecommand \BibitemOpen [0]{}%
\providecommand \bibitemStop [0]{}%
\providecommand \bibitemNoStop [0]{.\EOS\space}%
\providecommand \EOS [0]{\spacefactor3000\relax}%
\providecommand \BibitemShut  [1]{\csname bibitem#1\endcsname}%
\let\auto@bib@innerbib\@empty
\bibitem [{\citenamefont {Newman}(2003)}]{newman2003structure}%
  \BibitemOpen
  \bibfield  {author} {\bibinfo {author} {\bibfnamefont {M.~E.~J.}\
  \bibnamefont {Newman}},\ }\bibfield  {title} {\bibinfo {title} {The structure
  and function of complex networks},\ }\href
  {https://doi.org/10.1137/S003614450342480} {\bibfield  {journal} {\bibinfo
  {journal} {SIAM Review}\ }\textbf {\bibinfo {volume} {45}},\ \bibinfo {pages}
  {167} (\bibinfo {year} {2003})}\BibitemShut {NoStop}%
\bibitem [{\citenamefont {Arenas}\ \emph {et~al.}(2008)\citenamefont {Arenas},
  \citenamefont {D{\'\i}az-Guilera}, \citenamefont {Kurths}, \citenamefont
  {Moreno},\ and\ \citenamefont {Zhou}}]{arenas2008synchronization}%
  \BibitemOpen
  \bibfield  {author} {\bibinfo {author} {\bibfnamefont {A.}~\bibnamefont
  {Arenas}}, \bibinfo {author} {\bibfnamefont {A.}~\bibnamefont
  {D{\'\i}az-Guilera}}, \bibinfo {author} {\bibfnamefont {J.}~\bibnamefont
  {Kurths}}, \bibinfo {author} {\bibfnamefont {Y.}~\bibnamefont {Moreno}},\
  and\ \bibinfo {author} {\bibfnamefont {C.}~\bibnamefont {Zhou}},\ }\bibfield
  {title} {\bibinfo {title} {Synchronization in complex networks},\ }\href
  {https://doi.org/10.1016/j.physrep.2008.09.002} {\bibfield  {journal}
  {\bibinfo  {journal} {Physics Reports}\ }\textbf {\bibinfo {volume} {469}},\
  \bibinfo {pages} {93} (\bibinfo {year} {2008})}\BibitemShut {NoStop}%
\bibitem [{\citenamefont {Pagani}\ and\ \citenamefont
  {Aiello}(2013)}]{pagani2013power}%
  \BibitemOpen
  \bibfield  {author} {\bibinfo {author} {\bibfnamefont {G.~A.}\ \bibnamefont
  {Pagani}}\ and\ \bibinfo {author} {\bibfnamefont {M.}~\bibnamefont
  {Aiello}},\ }\bibfield  {title} {\bibinfo {title} {The power grid as a
  complex network: a survey},\ }\href
  {https://doi.org/10.1016/j.physa.2013.01.023} {\bibfield  {journal} {\bibinfo
   {journal} {Physica A: Statistical Mechanics and its Applications}\ }\textbf
  {\bibinfo {volume} {392}},\ \bibinfo {pages} {2688} (\bibinfo {year}
  {2013})}\BibitemShut {NoStop}%
\bibitem [{\citenamefont {Witthaut}\ \emph {et~al.}(2022)\citenamefont
  {Witthaut}, \citenamefont {Hellmann}, \citenamefont {Kurths}, \citenamefont
  {Kettemann}, \citenamefont {Meyer-Ortmanns},\ and\ \citenamefont
  {Timme}}]{witthaut2022collective}%
  \BibitemOpen
  \bibfield  {author} {\bibinfo {author} {\bibfnamefont {D.}~\bibnamefont
  {Witthaut}}, \bibinfo {author} {\bibfnamefont {F.}~\bibnamefont {Hellmann}},
  \bibinfo {author} {\bibfnamefont {J.}~\bibnamefont {Kurths}}, \bibinfo
  {author} {\bibfnamefont {S.}~\bibnamefont {Kettemann}}, \bibinfo {author}
  {\bibfnamefont {H.}~\bibnamefont {Meyer-Ortmanns}},\ and\ \bibinfo {author}
  {\bibfnamefont {M.}~\bibnamefont {Timme}},\ }\bibfield  {title} {\bibinfo
  {title} {Collective nonlinear dynamics and self-organization in decentralized
  power grids},\ }\href {https://doi.org/10.1103/RevModPhys.94.015005}
  {\bibfield  {journal} {\bibinfo  {journal} {Reviews of Modern Physics}\
  }\textbf {\bibinfo {volume} {94}},\ \bibinfo {pages} {015005} (\bibinfo
  {year} {2022})}\BibitemShut {NoStop}%
\bibitem [{\citenamefont {Newman}(2005)}]{newman2005power}%
  \BibitemOpen
  \bibfield  {author} {\bibinfo {author} {\bibfnamefont {M.~E.~J.}\
  \bibnamefont {Newman}},\ }\bibfield  {title} {\bibinfo {title} {Power laws,
  {P}areto distributions and {Z}ipf's law},\ }\href
  {https://doi.org/10.1080/00107510500052444} {\bibfield  {journal} {\bibinfo
  {journal} {Contemporary Physics}\ }\textbf {\bibinfo {volume} {46}},\
  \bibinfo {pages} {323} (\bibinfo {year} {2005})}\BibitemShut {NoStop}%
\bibitem [{\citenamefont {Clauset}\ \emph {et~al.}(2009)\citenamefont
  {Clauset}, \citenamefont {Shalizi},\ and\ \citenamefont
  {Newman}}]{Clauset2009}%
  \BibitemOpen
  \bibfield  {author} {\bibinfo {author} {\bibfnamefont {A.}~\bibnamefont
  {Clauset}}, \bibinfo {author} {\bibfnamefont {C.~R.}\ \bibnamefont
  {Shalizi}},\ and\ \bibinfo {author} {\bibfnamefont {M.~E.~J.}\ \bibnamefont
  {Newman}},\ }\bibfield  {title} {\bibinfo {title} {Power-law distributions in
  empirical data},\ }\href {https://doi.org/10.1137/070710111} {\bibfield
  {journal} {\bibinfo  {journal} {SIAM Review}\ }\textbf {\bibinfo {volume}
  {51}},\ \bibinfo {pages} {661} (\bibinfo {year} {2009})}\BibitemShut
  {NoStop}%
\bibitem [{\citenamefont {Albert}\ and\ \citenamefont
  {Barab{\'a}si}(2000)}]{albert2000topology}%
  \BibitemOpen
  \bibfield  {author} {\bibinfo {author} {\bibfnamefont {R.}~\bibnamefont
  {Albert}}\ and\ \bibinfo {author} {\bibfnamefont {A.-L.}\ \bibnamefont
  {Barab{\'a}si}},\ }\bibfield  {title} {\bibinfo {title} {Topology of evolving
  networks: local events and universality},\ }\href
  {https://doi.org/10.1103/PhysRevLett.85.5234} {\bibfield  {journal} {\bibinfo
   {journal} {Physical Review Letters}\ }\textbf {\bibinfo {volume} {85}},\
  \bibinfo {pages} {5234} (\bibinfo {year} {2000})}\BibitemShut {NoStop}%
\bibitem [{\citenamefont {Wu}\ \emph {et~al.}(2019)\citenamefont {Wu},
  \citenamefont {Chen}, \citenamefont {Yao}, \citenamefont {Zhao},\ and\
  \citenamefont {Chen}}]{wu2019fractal}%
  \BibitemOpen
  \bibfield  {author} {\bibinfo {author} {\bibfnamefont {Y.}~\bibnamefont
  {Wu}}, \bibinfo {author} {\bibfnamefont {Z.}~\bibnamefont {Chen}}, \bibinfo
  {author} {\bibfnamefont {K.}~\bibnamefont {Yao}}, \bibinfo {author}
  {\bibfnamefont {X.}~\bibnamefont {Zhao}},\ and\ \bibinfo {author}
  {\bibfnamefont {Y.}~\bibnamefont {Chen}},\ }\bibfield  {title} {\bibinfo
  {title} {On the correlation between fractal dimension and robustness of
  complex networks},\ }\href {https://doi.org/10.1142/S0218348X19500671}
  {\bibfield  {journal} {\bibinfo  {journal} {Fractals}\ }\textbf {\bibinfo
  {volume} {27}},\ \bibinfo {pages} {1950067} (\bibinfo {year}
  {2019})}\BibitemShut {NoStop}%
\bibitem [{\citenamefont {Anvari}\ \emph {et~al.}(2020)\citenamefont {Anvari},
  \citenamefont {Rydin~Gorj{\~a}o}, \citenamefont {Timme}, \citenamefont
  {Witthaut}, \citenamefont {Sch{\"a}fer},\ and\ \citenamefont
  {Kantz}}]{anvari2020stochastic}%
  \BibitemOpen
  \bibfield  {author} {\bibinfo {author} {\bibfnamefont {M.}~\bibnamefont
  {Anvari}}, \bibinfo {author} {\bibfnamefont {L.}~\bibnamefont
  {Rydin~Gorj{\~a}o}}, \bibinfo {author} {\bibfnamefont {M.}~\bibnamefont
  {Timme}}, \bibinfo {author} {\bibfnamefont {D.}~\bibnamefont {Witthaut}},
  \bibinfo {author} {\bibfnamefont {B.}~\bibnamefont {Sch{\"a}fer}},\ and\
  \bibinfo {author} {\bibfnamefont {H.}~\bibnamefont {Kantz}},\ }\bibfield
  {title} {\bibinfo {title} {Stochastic properties of the frequency dynamics in
  real and synthetic power grids},\ }\href
  {https://doi.org/10.1103/PhysRevResearch.2.013339} {\bibfield  {journal}
  {\bibinfo  {journal} {Physical Review Research}\ }\textbf {\bibinfo {volume}
  {2}},\ \bibinfo {pages} {013339} (\bibinfo {year} {2020})}\BibitemShut
  {NoStop}%
\bibitem [{\citenamefont {Machowski}\ \emph {et~al.}(2020)\citenamefont
  {Machowski}, \citenamefont {Lubosny}, \citenamefont {Bialek},\ and\
  \citenamefont {Bumby}}]{machowski2020power}%
  \BibitemOpen
  \bibfield  {author} {\bibinfo {author} {\bibfnamefont {J.}~\bibnamefont
  {Machowski}}, \bibinfo {author} {\bibfnamefont {Z.}~\bibnamefont {Lubosny}},
  \bibinfo {author} {\bibfnamefont {J.~W.}\ \bibnamefont {Bialek}},\ and\
  \bibinfo {author} {\bibfnamefont {J.~R.}\ \bibnamefont {Bumby}},\ }\href@noop
  {} {\emph {\bibinfo {title} {Power System Dynamics: Stability and
  Control}}},\ \bibinfo {edition} {3rd}\ ed.\ (\bibinfo  {publisher} {Wiley,
  Hoboken, New Jersey},\ \bibinfo {year} {2020})\BibitemShut {NoStop}%
\bibitem [{\citenamefont {Sch{\"a}fer}\ \emph {et~al.}(2018)\citenamefont
  {Sch{\"a}fer}, \citenamefont {Beck}, \citenamefont {Aihara}, \citenamefont
  {Witthaut},\ and\ \citenamefont {Timme}}]{schafer2018non}%
  \BibitemOpen
  \bibfield  {author} {\bibinfo {author} {\bibfnamefont {B.}~\bibnamefont
  {Sch{\"a}fer}}, \bibinfo {author} {\bibfnamefont {C.}~\bibnamefont {Beck}},
  \bibinfo {author} {\bibfnamefont {K.}~\bibnamefont {Aihara}}, \bibinfo
  {author} {\bibfnamefont {D.}~\bibnamefont {Witthaut}},\ and\ \bibinfo
  {author} {\bibfnamefont {M.}~\bibnamefont {Timme}},\ }\bibfield  {title}
  {\bibinfo {title} {Non-gaussian power grid frequency fluctuations
  characterized by {L}{\'e}vy-stable laws and superstatistics},\ }\href
  {https://doi.org/10.1038/s41560-017-0058-z} {\bibfield  {journal} {\bibinfo
  {journal} {Nature Energy}\ }\textbf {\bibinfo {volume} {3}},\ \bibinfo
  {pages} {119} (\bibinfo {year} {2018})}\BibitemShut {NoStop}%
\bibitem [{\citenamefont {Han}\ \emph {et~al.}(2022)\citenamefont {Han},
  \citenamefont {Witthaut}, \citenamefont {Rydin~Gorj{\~{a}}o},\ and\
  \citenamefont {Böttcher}}]{han2022collective}%
  \BibitemOpen
  \bibfield  {author} {\bibinfo {author} {\bibfnamefont {C.}~\bibnamefont
  {Han}}, \bibinfo {author} {\bibfnamefont {D.}~\bibnamefont {Witthaut}},
  \bibinfo {author} {\bibfnamefont {L.}~\bibnamefont {Rydin~Gorj{\~{a}}o}},\
  and\ \bibinfo {author} {\bibfnamefont {P.~C.}\ \bibnamefont {Böttcher}},\
  }\bibfield  {title} {\bibinfo {title} {Collective effects and synchronization
  of demand in real-time demand response},\ }\href
  {https://doi.org/10.1088/2632-072x/ac6477} {\bibfield  {journal} {\bibinfo
  {journal} {Journal of Physics: Complexity}\ }\textbf {\bibinfo {volume}
  {3}},\ \bibinfo {pages} {025002} (\bibinfo {year} {2022})}\BibitemShut
  {NoStop}%
\bibitem [{\citenamefont {Zhou}\ \emph {et~al.}(2020)\citenamefont {Zhou},
  \citenamefont {Lu}, \citenamefont {Li},\ and\ \citenamefont
  {Tan}}]{zhou2020fractal}%
  \BibitemOpen
  \bibfield  {author} {\bibinfo {author} {\bibfnamefont {T.}~\bibnamefont
  {Zhou}}, \bibinfo {author} {\bibfnamefont {J.}~\bibnamefont {Lu}}, \bibinfo
  {author} {\bibfnamefont {B.}~\bibnamefont {Li}},\ and\ \bibinfo {author}
  {\bibfnamefont {Y.}~\bibnamefont {Tan}},\ }\bibfield  {title} {\bibinfo
  {title} {Fractal analysis of power grid faults and cross correlation for the
  faults and meteorological factors},\ }\href
  {https://doi.org/10.1109/ACCESS.2020.2991109} {\bibfield  {journal} {\bibinfo
   {journal} {IEEE Access}\ }\textbf {\bibinfo {volume} {8}},\ \bibinfo {pages}
  {79935} (\bibinfo {year} {2020})}\BibitemShut {NoStop}%
\bibitem [{\citenamefont {Kruse}\ \emph {et~al.}(2021)\citenamefont {Kruse},
  \citenamefont {Sch{\"a}fer},\ and\ \citenamefont
  {Witthaut}}]{kruse2021revealing}%
  \BibitemOpen
  \bibfield  {author} {\bibinfo {author} {\bibfnamefont {J.}~\bibnamefont
  {Kruse}}, \bibinfo {author} {\bibfnamefont {B.}~\bibnamefont {Sch{\"a}fer}},\
  and\ \bibinfo {author} {\bibfnamefont {D.}~\bibnamefont {Witthaut}},\
  }\bibfield  {title} {\bibinfo {title} {Revealing drivers and risks for power
  grid frequency stability with explainable {AI}},\ }\href
  {https://doi.org/10.1016/j.patter.2021.100365} {\bibfield  {journal}
  {\bibinfo  {journal} {Patterns}\ }\textbf {\bibinfo {volume} {2}},\ \bibinfo
  {pages} {100365} (\bibinfo {year} {2021})}\BibitemShut {NoStop}%
\bibitem [{\citenamefont {Rydin~Gorj{\~a}o}\ \emph
  {et~al.}(2020{\natexlab{a}})\citenamefont {Rydin~Gorj{\~a}o}, \citenamefont
  {Jumar}, \citenamefont {Maa{\ss}}, \citenamefont {Hagenmeyer}, \citenamefont
  {Yalcin}, \citenamefont {Kruse}, \citenamefont {Timme}, \citenamefont {Beck},
  \citenamefont {Witthaut},\ and\ \citenamefont {Sch{\"a}fer}}]{rydin2020open}%
  \BibitemOpen
  \bibfield  {author} {\bibinfo {author} {\bibfnamefont {L.}~\bibnamefont
  {Rydin~Gorj{\~a}o}}, \bibinfo {author} {\bibfnamefont {R.}~\bibnamefont
  {Jumar}}, \bibinfo {author} {\bibfnamefont {H.}~\bibnamefont {Maa{\ss}}},
  \bibinfo {author} {\bibfnamefont {V.}~\bibnamefont {Hagenmeyer}}, \bibinfo
  {author} {\bibfnamefont {G.~C.}\ \bibnamefont {Yalcin}}, \bibinfo {author}
  {\bibfnamefont {J.}~\bibnamefont {Kruse}}, \bibinfo {author} {\bibfnamefont
  {M.}~\bibnamefont {Timme}}, \bibinfo {author} {\bibfnamefont
  {C.}~\bibnamefont {Beck}}, \bibinfo {author} {\bibfnamefont {D.}~\bibnamefont
  {Witthaut}},\ and\ \bibinfo {author} {\bibfnamefont {B.}~\bibnamefont
  {Sch{\"a}fer}},\ }\bibfield  {title} {\bibinfo {title} {Open database
  analysis of scaling and spatio-temporal properties of power grid
  frequencies},\ }\href {https://doi.org/10.1038/s41467-020-19732-7} {\bibfield
   {journal} {\bibinfo  {journal} {Nature Communications}\ }\textbf {\bibinfo
  {volume} {11}},\ \bibinfo {pages} {6362} (\bibinfo {year}
  {2020}{\natexlab{a}})}\BibitemShut {NoStop}%
\bibitem [{\citenamefont {{MagnaGen GmbH}}(2022)}]{Gridradar}%
  \BibitemOpen
  \bibfield  {author} {\bibinfo {author} {\bibnamefont {{MagnaGen GmbH}}},\
  }\href@noop {} {\bibinfo {title} {Gridradar--an independent grid monitoring
  system}},\ \bibinfo {howpublished} {\url{https://gridradar.net/}} (\bibinfo
  {year} {2022})\BibitemShut {NoStop}%
\bibitem [{\citenamefont {Chai}\ \emph {et~al.}(2016)\citenamefont {Chai},
  \citenamefont {Liu}, \citenamefont {Guo}, \citenamefont {Wu}, \citenamefont
  {Zhou}, \citenamefont {Yao}, \citenamefont {Liu}, \citenamefont {King},
  \citenamefont {Gracia},\ and\ \citenamefont {Patel}}]{chai_wide-area_2016}%
  \BibitemOpen
  \bibfield  {author} {\bibinfo {author} {\bibfnamefont {J.}~\bibnamefont
  {Chai}}, \bibinfo {author} {\bibfnamefont {Y.}~\bibnamefont {Liu}}, \bibinfo
  {author} {\bibfnamefont {J.}~\bibnamefont {Guo}}, \bibinfo {author}
  {\bibfnamefont {L.}~\bibnamefont {Wu}}, \bibinfo {author} {\bibfnamefont
  {D.}~\bibnamefont {Zhou}}, \bibinfo {author} {\bibfnamefont {W.}~\bibnamefont
  {Yao}}, \bibinfo {author} {\bibfnamefont {Y.}~\bibnamefont {Liu}}, \bibinfo
  {author} {\bibfnamefont {T.}~\bibnamefont {King}}, \bibinfo {author}
  {\bibfnamefont {J.~R.}\ \bibnamefont {Gracia}},\ and\ \bibinfo {author}
  {\bibfnamefont {M.}~\bibnamefont {Patel}},\ }\bibfield  {title} {\bibinfo
  {title} {Wide-area measurement data analytics using {FNET}/{GridEye}: {A}
  review},\ }in\ \href {https://doi.org/10.1109/PSCC.2016.7540946} {\emph
  {\bibinfo {booktitle} {2016 {Power} {Systems} {Computation} {Conference}
  ({PSCC})}}}\ (\bibinfo {year} {2016})\BibitemShut {NoStop}%
\bibitem [{\citenamefont {Maa{\ss}}\ \emph {et~al.}(2013)\citenamefont
  {Maa{\ss}}, \citenamefont {\c{C}akmak}, \citenamefont {Suess}, \citenamefont
  {Quinte}, \citenamefont {Jakob}, \citenamefont {Stucky},\ and\ \citenamefont
  {Kuehnapfel}}]{maass2013first}%
  \BibitemOpen
  \bibfield  {author} {\bibinfo {author} {\bibfnamefont {H.}~\bibnamefont
  {Maa{\ss}}}, \bibinfo {author} {\bibfnamefont {H.~K.}\ \bibnamefont
  {\c{C}akmak}}, \bibinfo {author} {\bibfnamefont {W.}~\bibnamefont {Suess}},
  \bibinfo {author} {\bibfnamefont {A.}~\bibnamefont {Quinte}}, \bibinfo
  {author} {\bibfnamefont {W.}~\bibnamefont {Jakob}}, \bibinfo {author}
  {\bibfnamefont {K.-U.}\ \bibnamefont {Stucky}},\ and\ \bibinfo {author}
  {\bibfnamefont {U.~G.}\ \bibnamefont {Kuehnapfel}},\ }\bibfield  {title}
  {\bibinfo {title} {First {Evaluation} {Results} {Using} the {New}
  {Electrical} {Data} {Recorder} for {Power} {Grid} {Analysis}},\ }\href
  {https://doi.org/10.1109/TIM.2013.2270923} {\bibfield  {journal} {\bibinfo
  {journal} {IEEE Transactions on Instrumentation and Measurement}\ }\textbf
  {\bibinfo {volume} {62}},\ \bibinfo {pages} {2384} (\bibinfo {year}
  {2013})}\BibitemShut {NoStop}%
\bibitem [{\citenamefont {Maa{\ss}}\ \emph {et~al.}(2015)\citenamefont
  {Maa{\ss}}, \citenamefont {Cakmak}, \citenamefont {Bach}, \citenamefont
  {Mikut}, \citenamefont {Harrabi}, \citenamefont {S{\"u}{\ss}}, \citenamefont
  {Jakob}, \citenamefont {Stucky}, \citenamefont {K{\"u}hnapfel},\ and\
  \citenamefont {Hagenmeyer}}]{maas2015data}%
  \BibitemOpen
  \bibfield  {author} {\bibinfo {author} {\bibfnamefont {H.}~\bibnamefont
  {Maa{\ss}}}, \bibinfo {author} {\bibfnamefont {H.~K.}\ \bibnamefont
  {Cakmak}}, \bibinfo {author} {\bibfnamefont {F.}~\bibnamefont {Bach}},
  \bibinfo {author} {\bibfnamefont {R.}~\bibnamefont {Mikut}}, \bibinfo
  {author} {\bibfnamefont {A.}~\bibnamefont {Harrabi}}, \bibinfo {author}
  {\bibfnamefont {W.}~\bibnamefont {S{\"u}{\ss}}}, \bibinfo {author}
  {\bibfnamefont {W.}~\bibnamefont {Jakob}}, \bibinfo {author} {\bibfnamefont
  {K.-U.}\ \bibnamefont {Stucky}}, \bibinfo {author} {\bibfnamefont {U.~G.}\
  \bibnamefont {K{\"u}hnapfel}},\ and\ \bibinfo {author} {\bibfnamefont
  {V.}~\bibnamefont {Hagenmeyer}},\ }\bibfield  {title} {\bibinfo {title}
  {{Data Processing of high-rate low-voltage Distribution Grid Recordings for
  Smart Grid Monitoring and Analysis}},\ }\href
  {https://doi.org/10.1186/s13634-015-0203-4} {\bibfield  {journal} {\bibinfo
  {journal} {EURASIP Journal on Advances in Signal Processing}\ }\textbf
  {\bibinfo {volume} {2015}},\ \bibinfo {pages} {14} (\bibinfo {year}
  {2015})}\BibitemShut {NoStop}%
\bibitem [{\citenamefont {Jumar}\ \emph {et~al.}(2021)\citenamefont {Jumar},
  \citenamefont {Maa{\ss}}, \citenamefont {Sch{\"a}fer}, \citenamefont
  {Rydin~Gorj{\~a}o},\ and\ \citenamefont {Hagenmeyer}}]{jumar2021database}%
  \BibitemOpen
  \bibfield  {author} {\bibinfo {author} {\bibfnamefont {R.}~\bibnamefont
  {Jumar}}, \bibinfo {author} {\bibfnamefont {H.}~\bibnamefont {Maa{\ss}}},
  \bibinfo {author} {\bibfnamefont {B.}~\bibnamefont {Sch{\"a}fer}}, \bibinfo
  {author} {\bibfnamefont {L.}~\bibnamefont {Rydin~Gorj{\~a}o}},\ and\ \bibinfo
  {author} {\bibfnamefont {V.}~\bibnamefont {Hagenmeyer}},\ }\bibfield  {title}
  {\bibinfo {title} {Database of {Power} {Grid} {Frequency} {Measurements}},\
  }\href {https://arxiv.org/abs/2006.01771} {\bibfield  {journal} {\bibinfo
  {journal} {arXiv preprint arXiv:2006.01771}\ } (\bibinfo {year}
  {2021})}\BibitemShut {NoStop}%
\bibitem [{\citenamefont {F{\"{o}}rstner}\ \emph {et~al.}(2022)\citenamefont
  {F{\"{o}}rstner}, \citenamefont {Jumar}, \citenamefont {Tanrikulu},
  \citenamefont {Maa{\ss }}, \citenamefont {K{\"{u}}hnapfel},\ and\
  \citenamefont {Hagenmeyer}}]{foerstner2022experimental}%
  \BibitemOpen
  \bibfield  {author} {\bibinfo {author} {\bibfnamefont {E.}~\bibnamefont
  {F{\"{o}}rstner}}, \bibinfo {author} {\bibfnamefont {R.}~\bibnamefont
  {Jumar}}, \bibinfo {author} {\bibfnamefont {O.~D.}\ \bibnamefont
  {Tanrikulu}}, \bibinfo {author} {\bibfnamefont {H.}~\bibnamefont {Maa{\ss
  }}}, \bibinfo {author} {\bibfnamefont {U.}~\bibnamefont {K{\"{u}}hnapfel}},\
  and\ \bibinfo {author} {\bibfnamefont {V.}~\bibnamefont {Hagenmeyer}},\
  }\bibfield  {title} {\bibinfo {title} {Experimental evaluation and
  systematic-error reduction of frequency estimation using the zero-crossing
  technique for dynamic power grids},\ }in\ \href
  {https://doi.org/10.1109/SGSMA51733.2022.9806021} {\emph {\bibinfo
  {booktitle} {2022 International Conference on Smart Grid Synchronized
  Measurements and Analytics (SGSMA)}}}\ (\bibinfo {year} {2022})\ pp.\
  \bibinfo {pages} {1--6}\BibitemShut {NoStop}%
\bibitem [{\citenamefont {Zhao}\ \emph {et~al.}(2020)\citenamefont {Zhao},
  \citenamefont {Zhan}, \citenamefont {Yin}, \citenamefont {Li}, \citenamefont
  {Yao},\ and\ \citenamefont {Liu}}]{zhao2020recent}%
  \BibitemOpen
  \bibfield  {author} {\bibinfo {author} {\bibfnamefont {J.}~\bibnamefont
  {Zhao}}, \bibinfo {author} {\bibfnamefont {L.}~\bibnamefont {Zhan}}, \bibinfo
  {author} {\bibfnamefont {H.}~\bibnamefont {Yin}}, \bibinfo {author}
  {\bibfnamefont {F.}~\bibnamefont {Li}}, \bibinfo {author} {\bibfnamefont
  {W.}~\bibnamefont {Yao}},\ and\ \bibinfo {author} {\bibfnamefont
  {Y.}~\bibnamefont {Liu}},\ }\bibfield  {title} {\bibinfo {title} {Recent
  development of frequency estimation methods for future smart grid},\ }\href
  {https://doi.org/10.1109/OAJPE.2020.3029984} {\bibfield  {journal} {\bibinfo
  {journal} {IEEE Open Access Journal of Power and Energy}\ }\textbf {\bibinfo
  {volume} {7}},\ \bibinfo {pages} {354} (\bibinfo {year} {2020})}\BibitemShut
  {NoStop}%
\bibitem [{\citenamefont {Grillo}\ \emph {et~al.}(2012)\citenamefont {Grillo},
  \citenamefont {Pasquino}, \citenamefont {Angrisani},\ and\ \citenamefont
  {Schiano Lo~Moriello}}]{grillo2012efficient}%
  \BibitemOpen
  \bibfield  {author} {\bibinfo {author} {\bibfnamefont {D.}~\bibnamefont
  {Grillo}}, \bibinfo {author} {\bibfnamefont {N.}~\bibnamefont {Pasquino}},
  \bibinfo {author} {\bibfnamefont {L.}~\bibnamefont {Angrisani}},\ and\
  \bibinfo {author} {\bibfnamefont {R.}~\bibnamefont {Schiano Lo~Moriello}},\
  }\bibfield  {title} {\bibinfo {title} {An efficient extension of the
  zero-crossing technique to measure frequency of noisy signals},\ }in\ \href
  {https://doi.org/10.1109/I2MTC.2012.6229703} {\emph {\bibinfo {booktitle}
  {2012 {IEEE} {International} {Instrumentation} and {Measurement} {Technology}
  {Conference} {Proceedings}}}}\ (\bibinfo {year} {2012})\ pp.\ \bibinfo
  {pages} {2706--2709}\BibitemShut {NoStop}%
\bibitem [{\citenamefont {Mendonça}\ \emph {et~al.}(2014)\citenamefont
  {Mendonça}, \citenamefont {Pinto},\ and\ \citenamefont
  {Duque}}]{mendonca2014least}%
  \BibitemOpen
  \bibfield  {author} {\bibinfo {author} {\bibfnamefont {T.~R.~F.}\
  \bibnamefont {Mendonça}}, \bibinfo {author} {\bibfnamefont {M.~F.}\
  \bibnamefont {Pinto}},\ and\ \bibinfo {author} {\bibfnamefont {C.~A.}\
  \bibnamefont {Duque}},\ }\bibfield  {title} {\bibinfo {title} {Least squares
  optimization of zero crossing technique for frequency estimation of power
  system grid distorted sinusoidal signals},\ }in\ \href
  {https://doi.org/10.1109/INDUSCON.2014.7059443} {\emph {\bibinfo {booktitle}
  {2014 11th {IEEE}/{IAS} {International} {Conference} on {Industry}
  {Applications}}}}\ (\bibinfo {year} {2014})\ pp.\ \bibinfo {pages}
  {1--6}\BibitemShut {NoStop}%
\bibitem [{\citenamefont {Luo}\ \emph {et~al.}(2018)\citenamefont {Luo},
  \citenamefont {Zhang}, \citenamefont {Bu}, \citenamefont {Zeng},
  \citenamefont {Zhang}, \citenamefont {Luo},\ and\ \citenamefont
  {Tang}}]{luo2018improved}%
  \BibitemOpen
  \bibfield  {author} {\bibinfo {author} {\bibfnamefont {X.}~\bibnamefont
  {Luo}}, \bibinfo {author} {\bibfnamefont {J.}~\bibnamefont {Zhang}}, \bibinfo
  {author} {\bibfnamefont {Q.}~\bibnamefont {Bu}}, \bibinfo {author}
  {\bibfnamefont {Q.}~\bibnamefont {Zeng}}, \bibinfo {author} {\bibfnamefont
  {Q.}~\bibnamefont {Zhang}}, \bibinfo {author} {\bibfnamefont
  {D.}~\bibnamefont {Luo}},\ and\ \bibinfo {author} {\bibfnamefont
  {L.}~\bibnamefont {Tang}},\ }\bibfield  {title} {\bibinfo {title} {Improved
  {Zero}-crossing {Method} for {Power} {System} {Frequency} {Estimation}},\
  }in\ \href {https://doi.org/10.1109/WCICA.2018.8630447} {\emph {\bibinfo
  {booktitle} {2018 13th {World} {Congress} on {Intelligent} {Control} and
  {Automation} ({WCICA})}}}\ (\bibinfo {year} {2018})\ pp.\ \bibinfo {pages}
  {1403--1407}\BibitemShut {NoStop}%
\bibitem [{\citenamefont {Rydin Gorj\~ao}\ \emph
  {et~al.}(2022{\natexlab{a}})\citenamefont {Rydin Gorj\~ao}, \citenamefont
  {Vanfretti}, \citenamefont {Witthaut}, \citenamefont {Beck},\ and\
  \citenamefont {Schäfer}}]{RydinGorjao2021a}%
  \BibitemOpen
  \bibfield  {author} {\bibinfo {author} {\bibfnamefont {L.}~\bibnamefont
  {Rydin Gorj\~ao}}, \bibinfo {author} {\bibfnamefont {L.}~\bibnamefont
  {Vanfretti}}, \bibinfo {author} {\bibfnamefont {D.}~\bibnamefont {Witthaut}},
  \bibinfo {author} {\bibfnamefont {C.}~\bibnamefont {Beck}},\ and\ \bibinfo
  {author} {\bibfnamefont {B.}~\bibnamefont {Schäfer}},\ }\bibfield  {title}
  {\bibinfo {title} {Phase and amplitude synchronization in power-grid
  frequency fluctuations in the nordic grid},\ }\href
  {https://doi.org/10.1109/ACCESS.2022.3150338} {\bibfield  {journal} {\bibinfo
   {journal} {IEEE Access}\ }\textbf {\bibinfo {volume} {10}},\ \bibinfo
  {pages} {18065} (\bibinfo {year} {2022}{\natexlab{a}})}\BibitemShut {NoStop}%
\bibitem [{\citenamefont {Rydin~Gorj{\~a}o}\ \emph {et~al.}(2021)\citenamefont
  {Rydin~Gorj{\~a}o}, \citenamefont {Sch{\"a}fer}, \citenamefont {Witthaut},\
  and\ \citenamefont {Beck}}]{RydinGorjao2021b}%
  \BibitemOpen
  \bibfield  {author} {\bibinfo {author} {\bibfnamefont {L.}~\bibnamefont
  {Rydin~Gorj{\~a}o}}, \bibinfo {author} {\bibfnamefont {B.}~\bibnamefont
  {Sch{\"a}fer}}, \bibinfo {author} {\bibfnamefont {D.}~\bibnamefont
  {Witthaut}},\ and\ \bibinfo {author} {\bibfnamefont {C.}~\bibnamefont
  {Beck}},\ }\bibfield  {title} {\bibinfo {title} {Spatio-temporal complexity
  of power-grid frequency fluctuations},\ }\href
  {https://doi.org/10.1088/1367-2630/ac08b3} {\bibfield  {journal} {\bibinfo
  {journal} {New Journal of Physics}\ }\textbf {\bibinfo {volume} {23}},\
  \bibinfo {pages} {073016} (\bibinfo {year} {2021})}\BibitemShut {NoStop}%
\bibitem [{\citenamefont {Guo}\ and\ \citenamefont
  {Schlipf}(2021)}]{spectral2021}%
  \BibitemOpen
  \bibfield  {author} {\bibinfo {author} {\bibfnamefont {F.}~\bibnamefont
  {Guo}}\ and\ \bibinfo {author} {\bibfnamefont {D.}~\bibnamefont {Schlipf}},\
  }\bibfield  {title} {\bibinfo {title} {A spectral model of grid frequency for
  assessing the impact of inertia response on wind turbine dynamics},\ }\href
  {https://doi.org/10.3390/en14092492} {\bibfield  {journal} {\bibinfo
  {journal} {Energies}\ }\textbf {\bibinfo {volume} {14}},\ \bibinfo {pages}
  {2492} (\bibinfo {year} {2021})}\BibitemShut {NoStop}%
\bibitem [{\citenamefont {Krakovsk{\'a}}\ and\ \citenamefont
  {Krakovsk{\'a}}(2016)}]{krakovska2016fractal}%
  \BibitemOpen
  \bibfield  {author} {\bibinfo {author} {\bibfnamefont {H.}~\bibnamefont
  {Krakovsk{\'a}}}\ and\ \bibinfo {author} {\bibfnamefont {A.}~\bibnamefont
  {Krakovsk{\'a}}},\ }\bibfield  {title} {\bibinfo {title} {Fractal dimension
  of self-affine signals: Four methods of estimation},\ }\href
  {https://doi.org/10.48550/arXiv.1611.06190} {\bibfield  {journal} {\bibinfo
  {journal} {arXiv preprint arXiv:1611.06190}\ } (\bibinfo {year}
  {2016})}\BibitemShut {NoStop}%
\bibitem [{\citenamefont {Mandelbrot}(1977)}]{mandelbrot1977}%
  \BibitemOpen
  \bibfield  {author} {\bibinfo {author} {\bibfnamefont {B.~B.}\ \bibnamefont
  {Mandelbrot}},\ }\href@noop {} {\emph {\bibinfo {title} {Fractals: Form,
  Chance, and Dimension}}},\ \bibinfo {edition} {1st}\ ed.\ (\bibinfo
  {publisher} {W. H. Freeman \& Company, San Francisco},\ \bibinfo {year}
  {1977})\BibitemShut {NoStop}%
\bibitem [{\citenamefont {Kang}\ \emph {et~al.}(2020)\citenamefont {Kang},
  \citenamefont {Kim}, \citenamefont {Mariani}, \citenamefont {Cho},
  \citenamefont {Winkelman}, \citenamefont {Park},\ and\ \citenamefont
  {Kang}}]{eeg2020}%
  \BibitemOpen
  \bibfield  {author} {\bibinfo {author} {\bibfnamefont {J.~M.}\ \bibnamefont
  {Kang}}, \bibinfo {author} {\bibfnamefont {S.~T.}\ \bibnamefont {Kim}},
  \bibinfo {author} {\bibfnamefont {S.}~\bibnamefont {Mariani}}, \bibinfo
  {author} {\bibfnamefont {S.-E.}\ \bibnamefont {Cho}}, \bibinfo {author}
  {\bibfnamefont {J.~W.}\ \bibnamefont {Winkelman}}, \bibinfo {author}
  {\bibfnamefont {K.~H.}\ \bibnamefont {Park}},\ and\ \bibinfo {author}
  {\bibfnamefont {S.-G.}\ \bibnamefont {Kang}},\ }\bibfield  {title} {\bibinfo
  {title} {Difference in spectral power density of sleep eeg between patients
  with simple snoring and those with obstructive sleep apnoea},\ }\href
  {https://doi.org/10.1038/s41598-020-62915-x} {\bibfield  {journal} {\bibinfo
  {journal} {Scientific Reports}\ }\textbf {\bibinfo {volume} {10}},\ \bibinfo
  {pages} {6135} (\bibinfo {year} {2020})}\BibitemShut {NoStop}%
\bibitem [{\citenamefont {Eke}\ \emph {et~al.}(2002)\citenamefont {Eke},
  \citenamefont {Herman}, \citenamefont {Kocsis},\ and\ \citenamefont
  {Kozak}}]{eke2002}%
  \BibitemOpen
  \bibfield  {author} {\bibinfo {author} {\bibfnamefont {A.}~\bibnamefont
  {Eke}}, \bibinfo {author} {\bibfnamefont {P.}~\bibnamefont {Herman}},
  \bibinfo {author} {\bibfnamefont {L.}~\bibnamefont {Kocsis}},\ and\ \bibinfo
  {author} {\bibfnamefont {L.~R.}\ \bibnamefont {Kozak}},\ }\bibfield  {title}
  {\bibinfo {title} {Fractal characterization of complexity in temporal
  physiological signals},\ }\href {https://doi.org/10.1088/0967-3334/23/1/201}
  {\bibfield  {journal} {\bibinfo  {journal} {Physiological Measurement}\
  }\textbf {\bibinfo {volume} {23}},\ \bibinfo {pages} {R1} (\bibinfo {year}
  {2002})}\BibitemShut {NoStop}%
\bibitem [{\citenamefont {Mandelbrot}(1983)}]{mandelbrot1983}%
  \BibitemOpen
  \bibfield  {author} {\bibinfo {author} {\bibfnamefont {B.~B.}\ \bibnamefont
  {Mandelbrot}},\ }\href@noop {} {\emph {\bibinfo {title} {The Fractal Geometry
  of Nature}}},\ \bibinfo {edition} {1st}\ ed.\ (\bibinfo  {publisher} {W. H.
  Freeman \& Company, San Francisco},\ \bibinfo {year} {1983})\BibitemShut
  {NoStop}%
\bibitem [{\citenamefont {Klein}\ \emph {et~al.}(1991)\citenamefont {Klein},
  \citenamefont {Rogers},\ and\ \citenamefont {Kundur}}]{Klein1991}%
  \BibitemOpen
  \bibfield  {author} {\bibinfo {author} {\bibfnamefont {M.}~\bibnamefont
  {Klein}}, \bibinfo {author} {\bibfnamefont {G.~J.}\ \bibnamefont {Rogers}},\
  and\ \bibinfo {author} {\bibfnamefont {P.}~\bibnamefont {Kundur}},\
  }\bibfield  {title} {\bibinfo {title} {A fundamental study of inter-area
  oscillations in power systems},\ }\href {https://doi.org/10.1109/59.119229}
  {\bibfield  {journal} {\bibinfo  {journal} {IEEE Transactions on Power
  Systems}\ }\textbf {\bibinfo {volume} {6}},\ \bibinfo {pages} {914} (\bibinfo
  {year} {1991})}\BibitemShut {NoStop}%
\bibitem [{\citenamefont {Pettersen}\ \emph {et~al.}(2014)\citenamefont
  {Pettersen}, \citenamefont {Lindén}, \citenamefont {Tetzlaff},\ and\
  \citenamefont {Einevoll}}]{bio2014}%
  \BibitemOpen
  \bibfield  {author} {\bibinfo {author} {\bibfnamefont {K.~H.}\ \bibnamefont
  {Pettersen}}, \bibinfo {author} {\bibfnamefont {H.}~\bibnamefont {Lindén}},
  \bibinfo {author} {\bibfnamefont {T.}~\bibnamefont {Tetzlaff}},\ and\
  \bibinfo {author} {\bibfnamefont {G.~T.}\ \bibnamefont {Einevoll}},\
  }\bibfield  {title} {\bibinfo {title} {Power laws from linear neuronal cable
  theory: {P}ower spectral densities of the soma potential, soma membrane
  current and single-neuron contribution to the {EEG}},\ }\href
  {https://doi.org/10.1371/journal.pcbi.1003928} {\bibfield  {journal}
  {\bibinfo  {journal} {PLOS Computational Biology}\ }\textbf {\bibinfo
  {volume} {10}},\ \bibinfo {pages} {1} (\bibinfo {year} {2014})}\BibitemShut
  {NoStop}%
\bibitem [{\citenamefont {Heide}\ \emph {et~al.}(2010)\citenamefont {Heide},
  \citenamefont {Von~Bremen}, \citenamefont {Greiner}, \citenamefont
  {Hoffmann}, \citenamefont {Speckmann},\ and\ \citenamefont
  {Bofinger}}]{heide2010seasonal}%
  \BibitemOpen
  \bibfield  {author} {\bibinfo {author} {\bibfnamefont {D.}~\bibnamefont
  {Heide}}, \bibinfo {author} {\bibfnamefont {L.}~\bibnamefont {Von~Bremen}},
  \bibinfo {author} {\bibfnamefont {M.}~\bibnamefont {Greiner}}, \bibinfo
  {author} {\bibfnamefont {C.}~\bibnamefont {Hoffmann}}, \bibinfo {author}
  {\bibfnamefont {M.}~\bibnamefont {Speckmann}},\ and\ \bibinfo {author}
  {\bibfnamefont {S.}~\bibnamefont {Bofinger}},\ }\bibfield  {title} {\bibinfo
  {title} {Seasonal optimal mix of wind and solar power in a future, highly
  renewable {E}urope},\ }\href {https://doi.org/10.1016/j.renene.2010.03.012}
  {\bibfield  {journal} {\bibinfo  {journal} {Renewable Energy}\ }\textbf
  {\bibinfo {volume} {35}},\ \bibinfo {pages} {2483} (\bibinfo {year}
  {2010})}\BibitemShut {NoStop}%
\bibitem [{\citenamefont {Ausloos}(2012)}]{ausloos2012generalized}%
  \BibitemOpen
  \bibfield  {author} {\bibinfo {author} {\bibfnamefont {M.}~\bibnamefont
  {Ausloos}},\ }\bibfield  {title} {\bibinfo {title} {Generalized {H}urst
  exponent and multifractal function of original and translated texts mapped
  into frequency and length time series},\ }\href
  {https://doi.org/10.1103/PhysRevE.86.031108} {\bibfield  {journal} {\bibinfo
  {journal} {Physical Review E}\ }\textbf {\bibinfo {volume} {86}},\ \bibinfo
  {pages} {031108} (\bibinfo {year} {2012})}\BibitemShut {NoStop}%
\bibitem [{\citenamefont {Hill}\ and\ \citenamefont
  {Bergen}(1982)}]{hill1982stability}%
  \BibitemOpen
  \bibfield  {author} {\bibinfo {author} {\bibfnamefont {D.}~\bibnamefont
  {Hill}}\ and\ \bibinfo {author} {\bibfnamefont {A.}~\bibnamefont {Bergen}},\
  }\bibfield  {title} {\bibinfo {title} {Stability analysis of multimachine
  power networks with linear frequency dependent loads},\ }\href@noop {}
  {\bibfield  {journal} {\bibinfo  {journal} {IEEE Transactions on Circuits and
  Systems}\ }\textbf {\bibinfo {volume} {29}},\ \bibinfo {pages} {840}
  (\bibinfo {year} {1982})}\BibitemShut {NoStop}%
\bibitem [{\citenamefont {Rydin Gorj\~ao}\ \emph
  {et~al.}(2022{\natexlab{b}})\citenamefont {Rydin Gorj\~ao}, \citenamefont
  {Hassan}, \citenamefont {Kurths},\ and\ \citenamefont
  {Witthaut}}]{RydinGorjao2022}%
  \BibitemOpen
  \bibfield  {author} {\bibinfo {author} {\bibfnamefont {L.}~\bibnamefont
  {Rydin Gorj\~ao}}, \bibinfo {author} {\bibfnamefont {G.}~\bibnamefont
  {Hassan}}, \bibinfo {author} {\bibfnamefont {J.}~\bibnamefont {Kurths}},\
  and\ \bibinfo {author} {\bibfnamefont {D.}~\bibnamefont {Witthaut}},\
  }\bibfield  {title} {\bibinfo {title} {{MFDFA}: {E}fficient multifractal
  detrended fluctuation analysis in python},\ }\href
  {https://doi.org/10.1016/j.cpc.2021.108254} {\bibfield  {journal} {\bibinfo
  {journal} {Computer Physics Communications}\ }\textbf {\bibinfo {volume}
  {273}},\ \bibinfo {pages} {108254} (\bibinfo {year}
  {2022}{\natexlab{b}})}\BibitemShut {NoStop}%
\bibitem [{\citenamefont {Beck}\ and\ \citenamefont
  {Cohen}(2003)}]{beck2003superstatistics}%
  \BibitemOpen
  \bibfield  {author} {\bibinfo {author} {\bibfnamefont {C.}~\bibnamefont
  {Beck}}\ and\ \bibinfo {author} {\bibfnamefont {E.~G.}\ \bibnamefont
  {Cohen}},\ }\bibfield  {title} {\bibinfo {title} {Superstatistics},\ }\href
  {https://doi.org/10.1016/S0378-4371(03)00019-0} {\bibfield  {journal}
  {\bibinfo  {journal} {Physica A: Statistical mechanics and its applications}\
  }\textbf {\bibinfo {volume} {322}},\ \bibinfo {pages} {267} (\bibinfo {year}
  {2003})}\BibitemShut {NoStop}%
\bibitem [{\citenamefont {Friedrich}\ \emph {et~al.}(2022)\citenamefont
  {Friedrich}, \citenamefont {Moreno}, \citenamefont {Sinhuber}, \citenamefont
  {Wächter},\ and\ \citenamefont {Peinke}}]{Friedrich2022}%
  \BibitemOpen
  \bibfield  {author} {\bibinfo {author} {\bibfnamefont {J.}~\bibnamefont
  {Friedrich}}, \bibinfo {author} {\bibfnamefont {D.}~\bibnamefont {Moreno}},
  \bibinfo {author} {\bibfnamefont {M.}~\bibnamefont {Sinhuber}}, \bibinfo
  {author} {\bibfnamefont {M.}~\bibnamefont {Wächter}},\ and\ \bibinfo
  {author} {\bibfnamefont {J.}~\bibnamefont {Peinke}},\ }\bibfield  {title}
  {\bibinfo {title} {Surrogate modelling of wind fields from point-wise
  atmospheric turbulence measurements},\ }\href
  {https://doi.org/10.1088/1742-6596/2265/2/022026} {\bibfield  {journal}
  {\bibinfo  {journal} {Journal of Physics: Conference Series}\ }\textbf
  {\bibinfo {volume} {2265}},\ \bibinfo {pages} {022026} (\bibinfo {year}
  {2022})}\BibitemShut {NoStop}%
\bibitem [{\citenamefont {Rydin~Gorj{\~a}o}\ \emph
  {et~al.}(2020{\natexlab{b}})\citenamefont {Rydin~Gorj{\~a}o}, \citenamefont
  {Anvari}, \citenamefont {Kantz}, \citenamefont {Beck}, \citenamefont
  {Witthaut}, \citenamefont {Timme},\ and\ \citenamefont
  {Sch{\"a}fer}}]{gorjao2020data}%
  \BibitemOpen
  \bibfield  {author} {\bibinfo {author} {\bibfnamefont {L.}~\bibnamefont
  {Rydin~Gorj{\~a}o}}, \bibinfo {author} {\bibfnamefont {M.}~\bibnamefont
  {Anvari}}, \bibinfo {author} {\bibfnamefont {H.}~\bibnamefont {Kantz}},
  \bibinfo {author} {\bibfnamefont {C.}~\bibnamefont {Beck}}, \bibinfo {author}
  {\bibfnamefont {D.}~\bibnamefont {Witthaut}}, \bibinfo {author}
  {\bibfnamefont {M.}~\bibnamefont {Timme}},\ and\ \bibinfo {author}
  {\bibfnamefont {B.}~\bibnamefont {Sch{\"a}fer}},\ }\bibfield  {title}
  {\bibinfo {title} {Data-driven model of the power-grid frequency dynamics},\
  }\href {https://doi.org/10.1109/ACCESS.2020.2967834} {\bibfield  {journal}
  {\bibinfo  {journal} {IEEE Access}\ }\textbf {\bibinfo {volume} {8}},\
  \bibinfo {pages} {43082} (\bibinfo {year} {2020}{\natexlab{b}})}\BibitemShut
  {NoStop}%
\bibitem [{\citenamefont {Weixelbraun}\ \emph {et~al.}(2009)\citenamefont
  {Weixelbraun}, \citenamefont {Renner}, \citenamefont {Schmaranz},\ and\
  \citenamefont {Marketz}}]{weixelbraun2009dynamic}%
  \BibitemOpen
  \bibfield  {author} {\bibinfo {author} {\bibfnamefont {M.}~\bibnamefont
  {Weixelbraun}}, \bibinfo {author} {\bibfnamefont {H.}~\bibnamefont {Renner}},
  \bibinfo {author} {\bibfnamefont {R.}~\bibnamefont {Schmaranz}},\ and\
  \bibinfo {author} {\bibfnamefont {M.}~\bibnamefont {Marketz}},\ }\bibfield
  {title} {\bibinfo {title} {Dynamic simulation of a 110-k{V}-network during
  grid restoration and in islanded operation},\ }in\ \href
  {https://doi.org/10.1049/cp.2009.0740} {\emph {\bibinfo {booktitle} {CIRED
  2009-20th International Conference and Exhibition on Electricity
  Distribution-Part 1}}}\ (\bibinfo {organization} {IET},\ \bibinfo {year}
  {2009})\ pp.\ \bibinfo {pages} {1--4}\BibitemShut {NoStop}%
\bibitem [{\citenamefont {{Maa{\ss}}}\ \emph {et~al.}(2014)\citenamefont
  {{Maa{\ss}}}, \citenamefont {{Çakmak}}, \citenamefont {{Bach}},\ and\
  \citenamefont {{Kühnapfel}}}]{EDRprep2014}%
  \BibitemOpen
  \bibfield  {author} {\bibinfo {author} {\bibfnamefont {H.}~\bibnamefont
  {{Maa{\ss}}}}, \bibinfo {author} {\bibfnamefont {H.~K.}\ \bibnamefont
  {{Çakmak}}}, \bibinfo {author} {\bibfnamefont {F.}~\bibnamefont {{Bach}}},\
  and\ \bibinfo {author} {\bibfnamefont {U.~G.}\ \bibnamefont {{Kühnapfel}}},\
  }\bibfield  {title} {\bibinfo {title} {Preparing the electrical data recorder
  for comparative power network measurements},\ }in\ \href
  {https://doi.org/10.1109/ENERGYCON.2014.6850511} {\emph {\bibinfo {booktitle}
  {2014 IEEE International Energy Conference (ENERGYCON)}}}\ (\bibinfo {year}
  {2014})\ pp.\ \bibinfo {pages} {759--765}\BibitemShut {NoStop}%
\bibitem [{\citenamefont {{Jumar}}\ \emph {et~al.}(2019)\citenamefont
  {{Jumar}}, \citenamefont {{Maaß}}, \citenamefont {{Kühnapfel}},\ and\
  \citenamefont {{Hagenmeyer}}}]{EDR_Rework2019}%
  \BibitemOpen
  \bibfield  {author} {\bibinfo {author} {\bibfnamefont {R.}~\bibnamefont
  {{Jumar}}}, \bibinfo {author} {\bibfnamefont {H.}~\bibnamefont {{Maaß}}},
  \bibinfo {author} {\bibfnamefont {U.}~\bibnamefont {{Kühnapfel}}},\ and\
  \bibinfo {author} {\bibfnamefont {V.}~\bibnamefont {{Hagenmeyer}}},\
  }\bibfield  {title} {\bibinfo {title} {Synchronized continuous high-rate
  time-series recording in distribution grids for accurate evaluation},\ }in\
  \href {https://doi.org/10.1109/SGSMA.2019.8784664} {\emph {\bibinfo
  {booktitle} {2019 International Conference on Smart Grid Synchronized
  Measurements and Analytics (SGSMA)}}}\ (\bibinfo {year} {2019})\ pp.\
  \bibinfo {pages} {1--8}\BibitemShut {NoStop}%
\bibitem [{\citenamefont {Proakis}(2013)}]{Proakis2013}%
  \BibitemOpen
  \bibfield  {author} {\bibinfo {author} {\bibfnamefont {J.~G.}\ \bibnamefont
  {Proakis}},\ }\href@noop {} {\emph {\bibinfo {title} {Fundamentals of
  Communication Systems}}},\ \bibinfo {edition} {2nd}\ ed.\ (\bibinfo
  {publisher} {Pearson},\ \bibinfo {address} {Boston},\ \bibinfo {year}
  {2013})\BibitemShut {NoStop}%
\bibitem [{\citenamefont {Tyloo}(2022)}]{tyloo2022layered}%
  \BibitemOpen
  \bibfield  {author} {\bibinfo {author} {\bibfnamefont {M.}~\bibnamefont
  {Tyloo}},\ }\bibfield  {title} {\bibinfo {title} {Layered complex networks as
  fluctuation amplifiers},\ }\href@noop {} {\bibfield  {journal} {\bibinfo
  {journal} {arXiv preprint arXiv:2204.10251}\ } (\bibinfo {year}
  {2022})}\BibitemShut {NoStop}%
\bibitem [{\citenamefont {Sim}\ \emph {et~al.}(2022)\citenamefont {Sim},
  \citenamefont {Peinke},\ and\ \citenamefont {Maass}}]{sim2022offshore}%
  \BibitemOpen
  \bibfield  {author} {\bibinfo {author} {\bibfnamefont {S.-K.}\ \bibnamefont
  {Sim}}, \bibinfo {author} {\bibfnamefont {J.}~\bibnamefont {Peinke}},\ and\
  \bibinfo {author} {\bibfnamefont {P.}~\bibnamefont {Maass}},\ }\bibfield
  {title} {\bibinfo {title} {Offshore wind: {E}vidence for two-dimensional
  turbulence and role of sea horizon},\ }\href
  {https://doi.org/10.48550/arXiv.2203.07685} {\bibfield  {journal} {\bibinfo
  {journal} {arXiv preprint arXiv:2203.07685}\ } (\bibinfo {year}
  {2022})}\BibitemShut {NoStop}%
\bibitem [{\citenamefont {authors}(2022)}]{github}%
  \BibitemOpen
  \bibfield  {author} {\bibinfo {author} {\bibfnamefont {T.}~\bibnamefont
  {authors}},\ }\href@noop {} {\bibinfo {title} {Github repository with code to
  reproduce key results.}},\ \bibinfo {howpublished} {Link will be included
  once accepted} (\bibinfo {year} {2022})\BibitemShut {NoStop}%
\end{thebibliography}%

\end{document}